\documentclass[aps,twocolumn,pra,superscriptaddress]{revtex4-2}
\usepackage{amsfonts}
\usepackage[final,hiresbb]{graphicx}
\usepackage{amsmath}
\usepackage{amssymb}
\usepackage{amstext}
\usepackage{color}
\usepackage{braket}
\usepackage{leftindex}

\usepackage{subfigure}
\usepackage{appendix}
\usepackage{bbm}
\usepackage{bbold}

\usepackage{enumitem}

\definecolor{Green}{RGB}{0,204,102}
\definecolor{Purple}{RGB}{102,0,255}
\definecolor{Blue}{RGB}{51,153,255}
\definecolor{Red}{RGB}{255,010,010}

\newcommand{\rsr}{r_{\!s\!/\!r}}
\newcommand{\rTP}{r_{\!tp}}
\newcommand{\Cplus}{{\cal C}_+}

\newcommand{\smax}{{s}_{\scriptstyle\rm max}}
\newcommand{\GFR}{\Phi_{\!\scriptstyle\rm GFR}}
\newcommand{\APX}{{\bf A}^{\!\scriptstyle\rm (PX3)}}

\newcommand{\PPX}{{\bf P}_A^{\!\scriptstyle\rm (PX3)}}
\newcommand{\PFW}{{\bf P}_A^{\!\scriptstyle\rm (FW3)}}

\DeclareMathOperator{\sign}{sgn}

\begin{document}
	\title{Gravitational Faraday Holonomy}
	
\author{Blake A. Parvin}
\affiliation{Department of Physics, Colorado School of Mines, Golden, CO 80401, USA}

\author{Mark T. Lusk}
\email{mlusk@mines.edu}
\affiliation{Department of Physics, Colorado School of Mines, Golden, CO 80401, USA}

\begin{abstract}
Closed optical trajectories in Kerr spacetime are engineered to exhibit a marked lack of symmetry. The eccentricity manifests as a holonomy in gravitational Faraday rotation that can be made arbitrarily large by radial translation of the common location of source and receiver. All trajectories are non-equatorial and include a passage through the equatorial plane at the radial turning point, where the trajectory and pseudo-magnetic field are well-aligned. This, combined with path asymmetry, results in a large gravitational Faraday holonomy that lends itself to experimental measurement. Trajectories that start further away from the singularity pass more closely to the ergosphere, thus transiting  a more distorted region of spacetime with concomitant amplification of gravitational Coriolis force.
\end{abstract}
\maketitle

\section{Introduction}

The path of light in curved spacetime is well-approximated as a geodesic with a polarization that is parallel transported (PX). Rotation of the plane of polarization is the minimum required to maintain orthogonality to the path, so it also Fermi-Walker transported (FWX)\cite{ThorneMisnerWheeler}. The projection of both path and polarization into three-space, though, gives the perception that the latter rotates about the travel axis in excess of the minimum prescription of three-space Fermi-Walker transport. The additional gyration, a Coriolis effect that can be framed in terms of a fictitious magnetic field, is referred to as gravitational Faraday rotation (GFR). It has been theoretically considered for decades\cite{Plebanski_1960, ThorneMisnerWheeler, FrolovShoom2011, Oancea_2020}, but the effect has yet to be experimentally measured.

The appraisal of GFR would be simplified if light could be induced to traverse a closed circuit so that emission and reception could be carried out at a common location. Since the rotation is zero at the time of emission, measurement of a finite GFR at the receiver would amount to a gravitational Faraday holonomy (GFH). As a first step in this direction, a class of closed ambits has recently been considered in association with Kerr spacetime\cite{Lusk_PRD_2024}, characterized by travel with a constant Boyer-Lindquist radius. The outgoing and incoming directions of travel are not collinear, and the polarizations are distinct as well. However, the GFR increases and decreases in equal proportions along such paths, consistent with their highly symmetric, clover-leaf appearance, so there is no Faraday holonomy. This motivates the search for closed trajectories that exhibit three-space asymmetry, imbalancing the rise and fall of GFR and resulting in an experimentally measurable gravitational Faraday holonomy. 

In the present work, we identify a new type of non-equatorial, asymmetric, closed trajectory. These have a single radial turning point midway, where they dive through the equatorial plane near to the ergosphere horizon. The paths comprise an open interval in a parameter space of scalar Killing invariants. It is bounded at one end by a trajectory that enters the ergosphere and at the other by a trajectory that is iso-radial so that there is no holonomy. Trajectories associated with interior parameter points are shown to exhibit a holonomy in gravitational Faraday rotation that can be expressed analytically.  The holonomy increases as the source/receiver is moved outwards within the equatorial plane, and it can be engineered to be arbitrarily large. This is explained using numerical simulations that make it possible to visualize the difference between three-space Fermi-Walker transport and the perceived evolution of polarization.

\section{Approach}
  	
\subsection{Propagation of Light and Polarization within a Geometric Optics Approximation}

Starting with the source-free Maxwell equation in a spacetime with metric $g_{\mu\nu}$~\cite{ThorneMisnerWheeler}, an Eikonal expansion of the vector potential can be carried out
in terms of the perturbation ratio of wavelength over characteristic length\cite{Lusk_PRD_2024}. The wave vector, ${\bf p}$, then has null character at leading order in the expansion:
\begin{equation}\label{pnull}
{\bf p} \cdot {\bf p} = 0.
\end{equation}
Of course, this is equivalent to saying that the wave vector is parallel transported, 
\begin{equation}\label{pPX}
p^\nu \nabla_\nu p_\mu = 0,
\end{equation}
a geodesic equation that governs the trajectory of light\cite{ThorneMisnerWheeler}. The Eikonal expansion also implies that polarization, ${\bf f}$, is orthogonal to the trajectory at leading order:
\begin{equation}\label{orthogpol}
{\bf f} \cdot {\bf p} = 0.
\end{equation}
It is therefore parallel transported as well:
\begin{equation}\label{fPX}
p^\nu \nabla_\nu f_\mu = 0.
\end{equation}

In this work, attention is restricted to spacetime with a Kerr metric, expressed using oblate spheroidal Boyer-Lindquist coordinates\cite{BoyerLindquist1967}, $\{t, r,\theta,\phi\}$, with covariant components of
\begin{equation}\label{KerrBL}
[g] = \left(
\begin{array}{cccc}
 \frac{2 r}{\Sigma }-1 & 0 & 0 & -\frac{2 a r  \mathbb{s}^2 }{\Sigma } \\
 0 & \frac{\Sigma }{\Delta } & 0 & 0 \\
 0 & 0 & \Sigma  & 0 \\
-\frac{2 a r  \mathbb{s}^2 }{\Sigma }  & 0 & 0 &  \frac{\mathbb{s}^2}{\Sigma} \left( (a^2+r^2)^2   - \mathbb{s}^2 \Delta a^2 \right) \\
\end{array}
\right).
\end{equation}
Here $ \mathbb{c} := \cos\theta, \quad \mathbb{s}  := \sin\theta$, $\Sigma := r^2 + a^2 \mathbb{c}^2$, $\Delta := a^2 - 2 r + r^2$, and $a$ is the angular momentum of the black hole. Geometrized units are the default unless otherwise noted.

Isometries associated with this metric make it possible to express the evolution equation for trajectory in terms of three conserved scalars. These, in turn, can be solved in terms of Jacobi elliptic functions. The polarization along such paths can then be determined in closed form by exploiting a hidden symmetry associated with a conformal Killing-Yano tensor.

Metric isometries are associated with Lie derivatives of the metric that are zero with respect to Killing vectors, $\bf v$, or Killing tensors, $\hat K$\cite{Carter1968, Walker_Penrose_1970, Chandrasekhar1998, Carroll2004, WiltshireVisser2009}. These can be written as
\begin{equation}\label{Killing2}
\nabla_{(\mu} v_{\nu)}  = 0  , \quad \nabla_{(\mu} K_{\nu \gamma)} = 0,
\end{equation}
where parentheses indicate the symmetry operator. Since neither time, $t$, nor azimuthal angle, $\phi$, appear in the Kerr metric, it has two Killing vectors, ${\bf v}_t$ and ${\bf v}_\phi$, with Boyer-Lindquist coordinates of $\{1,0,0,0\}$ and $\{0,0,0,1\}$, respectively. The associated conserved scalars are
\begin{equation}\label{Killing_Vectors}
\epsilon := -{\bf v}_t \cdot {\bf p}, \quad \ell := {\bf v}_\phi \cdot {\bf p}.
\end{equation}
Noting that $dt/d\tau \equiv p^0$ and $d\phi/d\tau \equiv p^3$, for affine parameter $\tau$, these are equivalent to ordinary differential equations (ODEs) for the evolution of polarization components: 
\begin{align}\label{Killing_Vectors_2}
\varepsilon &= \frac{2 a p^3 r  \mathbb{s}^2}{\Sigma
   }-p^0\left(\frac{2 r}{\Sigma } - 1\right)  \nonumber \\
\ell &= \frac{ \mathbb{s}^2 \left(p^3
   \left(\left(a^2+r^2\right)^2 - a^2 \Delta  \mathbb{s}^2\right)-2 a p^0 r\right)}{\Sigma
   }.
  \end{align}

The Killing tensor of Eq. \ref{Killing2} has matrix representation
\begin{equation}\label{Ktensor}
[K]=\left(
\begin{array}{cccc}
 a^2 \left(1-\frac{2 \mathbb{c} ^2 r}{\Sigma }\right) & 0 &
   0 & K_{t \phi} \\
 0 & -\frac{a^2 \mathbb{c} ^2 \Sigma }{\Delta } & 0 & 0 \\
 0 & 0 & r^2 \Sigma  & 0 \\
K_{t \phi} & 0 & 0 &
K_{\phi\phi} \\
\end{array}
\right),
\end{equation}
where
\begin{align}
&K_{t \phi} =   -\frac{a \mathbb{s} ^2 \left(a^2 \mathbb{c} ^2 \Delta +r^2
   \left(a^2+r^2\right)\right)}{\Sigma }\\
&K_{\phi\phi} =  \frac{\mathbb{s} ^2 \left(a^4 \Delta  \mathbb{c} ^2 \mathbb{s} ^2+r^2
   \left(a^2+r^2\right)^2\right)}{\Sigma } .
\end{align}
%
 %
Its conserved scalar is $|\mathbb{k}|$, so that
\begin{equation}\label{kmag}
 |\mathbb{k}| = {\bf p} \cdot \hat K \cdot {\bf p}.
\end{equation}
Here\cite{Connors_1977, Chandrasekhar1998}
\begin{equation}\label{k_Kerr}
\mathbb{k} = r \alpha -a \beta \mathbb{c}  -  \imath  ( r \beta + a \alpha  \mathbb{c} ),
\end{equation}
with
\begin{align}\label{k_Kerr_2}
&\alpha := f^1 p^0 + a (f^3 p^1 - f^1 p^3) \mathbb{s} ^2 \nonumber \\
&\beta := -a f^2 p^0  \mathbb{s}  + (-f^3 p^2 + f^2 p^3) (a^2 + r^2)  \mathbb{s} .
\end{align}
%

%
The conserved scalar, $|\mathbb{k}|$, can be expressed in terms of Carter's constant, $Q$\cite{Carter1968, Chandrasekhar1998}:
\begin{equation}\label{Q}
 |\mathbb{k}|  = Q + (\ell - a\varepsilon)^2  .
\end{equation}

Eq. \ref{kmag} is therefore a first-order ODE involving the rate of change  $dr/d\tau \equiv p^1$:
\begin{equation}\label{kmag2}
 Q + (\ell - a\varepsilon)^2 = \frac{\left(\varepsilon  \left(a^2+r^2\right)-a \ell
   \right)^2}{\Delta }-\frac{(p^1)^2 \Sigma
   ^2}{\Delta }.
\end{equation}
%

Finally, the Boyer-Lindquist expression for Eq. \ref{pnull} amounts to a first-order ODE in terms of both $p^1$ and $d\theta/d\tau \equiv p^2$:
\begin{align}\label{pnull2}
4 \varepsilon^2 r (a^2+r^2)&+2 a^2 \ell
   ^2 -8 a \varepsilon  r \ell +2 \Delta  \varepsilon
   ^2 \Sigma \nonumber \\
= \, \ell^2 ( a^2+\Delta )& \csc ^2\theta + 2
   \Sigma^2 ((p^1)^2 + (p^2)^2 \Delta ).
\end{align}

Eqs. \ref{Killing_Vectors_2}$_{a,b}$, \ref{kmag2}, and \ref{pnull2} comprise four first-order ODEs for the four spacetime trajectory components\cite{StewartandWalker1973}. They are linear in the derivative terms and so can easily be re-arranged as separate ODEs for each position coordinate. In that form, though, they are still coupled because each derivative carries a coefficient of $\Sigma = r^2 + a^2 \mathbb{c} $. This can be removed with a simple re-scaling of parametrization from $\tau$ to  Mino time, $s$\cite{Mino_2003}:
\begin{equation}\label{Mino}
\frac{dx^\mu}{ds} := \frac{\Sigma}{\epsilon} p^\mu \equiv \frac{\Sigma}{\epsilon}  \frac{dx^\mu}{d\tau} .
\end{equation}

It is also useful to observe that there are only two independent scalars in these equations, impact parameter $\lambda := \ell / \varepsilon$ and Carter ratio $\eta:= Q / \varepsilon^2$. The well-known resulting equations of motion are
\begin{align}
\left(\frac{dr}{ds}\right)^2 &= {\mathcal R}(r) \label{EoMr}\\
\left(\frac{d\theta}{ds}\right)^2 &= \Theta(\theta) \label{EoMtheta}\\
\frac{d\phi}{ds} &= \frac{a}{\Delta}(r^2+a^2 - a \lambda) + \frac{\lambda}{\mathbb{s}^2} - a \label{EoMphi}\\
\frac{d t}{ds} &= \frac{(r^2+a^2)}{\Delta}(r^2+a^2 - a \lambda) + a(\lambda - a \mathbb{s}^2), \label{EoMt}
\end{align}
where radial potential, ${\mathcal R}(r)$, and polar potential, ${\Theta}(\theta)$, are
\begin{align}
{\mathcal R}(r) &:=  \left(a^2-a \lambda
   +r^2\right)^2-\Delta 
   \left((\lambda -a)^2+\eta
   \right) \label{Rpotential}\\
\Theta(\theta) &:= \eta - (a \mathbb{c})^2 - \left( \lambda \mathbb{c}/\mathbb{s}\right)^2 .\label{Thetapotential}
\end{align}
Given a position for the light source, and characterizing its direction with parameters $\lambda$ and $\eta$, Eqs. \ref{EoMr}  and \ref{EoMtheta} can be solved independently to obtain the radial and polar trajectories. The results can then be substituted into the Eqs. \ref{EoMphi}  and \ref{EoMt} to obtain the azimuthal and temporal trajectories.

\subsection{Non-Equatorial Asymmetric Trajectories}

The trajectories of interest in the present work have solutions to the equations of motion in terms of Jacobi elliptic functions: 
\begin{align}\label{traj1}
r(s) &= \frac{r_4 (r_3 - r_1) - r_3(r_4 - r_1){\rm sn}^2(X_2(s), k)}{(r_3 - r_1) - (r_4 - r_1){\rm sn}^2(X_2(s), k) } \nonumber \\
\theta(s) &= \cos^{-1} \left[ -\nu_\theta \sqrt{u_+} \,\,{\rm sn} \left(  \sqrt{-a^2 u_-} (s + \nu_\theta \mathcal{G}_\theta ), \frac{u_+}{u_-} \right) \right] \nonumber \\
\phi(s) &= \lambda G_\phi  \nonumber\\
 + &\frac{2 a}{r_+ - r_-} \left[ \left( r_+ - \frac{a \lambda}{2} \right) I_+(s) - \left( r_- - \frac{a \lambda}{2} \right) I_-(s) \right] \nonumber \\
t(s) &= I(s) + a^2 G(s) . 
\end{align}
These are identical to results derived elsewhere\cite{Gralla_2020} except for small corrections to the expressions for $\phi(s)$ and $t(s)$. All requisite functions are defined in Appendix A, where notes are also provided on the corrections.

\subsection{Parallel Transport of Polarization in Spacetime}

With analytical trajectory constructions now available, we turn to the evolution of polarization. Here it is useful to draw on symmetries that are not associated with configurational isometries but, rather, symmetry operations in the dynamical state space. These are referred to as \emph{hidden symmetries}, and the associated operations are manifested in a generalization of Eq. \ref{Killing2}(b) to the conformal Killing-Yano equation\cite{Frolov_LRR_2017}:
\begin{equation}\label{KYeqn}
\nabla_{\mu} H_{\nu\lambda}  = \frac{1}{3}g_{\mu\nu}  \nabla^\gamma H_{\gamma\lambda}  -  \frac{1}{3}g_{\mu\lambda}  \nabla^\gamma H_{\gamma\nu} .
\end{equation}
The skew-symmetric 2-form, $\hat H$, is the \emph{Principal Tensor}\cite{Frolov_LRR_2017}, and its Hodge dual, $\hat F := \null^*\!\hat H$, is also a conformal Killing-Yano tensor. These fields are considered more fundamental than Killing tensors since their square always generates a symmetric Killing tensor,
\begin{equation}\label{KYprop}
K_{\mu\nu}  = F_{\mu\gamma} F_{\nu\alpha}  g^{\alpha\gamma},
\end{equation}
while the reverse is not necessarily true.

In Boyer-Lindquist coordinates, the covariant components are:
\begin{equation}\label{Htensor}
[H]=\left(
\begin{array}{cccc}
 0 & r & a^2 \mathbb{c} \mathbb{s} & 0 \\
 -r & 0 & 0 & a r \mathbb{s} ^2 \\
 -a^2 \mathbb{c} \mathbb{s} & 0 & 0 & a \mathbb{c} \mathbb{s} \left(a^2+r^2\right) \\
 0 & -a r \mathbb{s} ^2 & -a \mathbb{c} \mathbb{s} \left(a^2+r^2\right) & 0 \\
\end{array}
\right)
\end{equation}
and
\begin{equation}\label{Ftensor}
[F]=\left(
\begin{array}{cccc}
 0 & -a \mathbb{c} & a r \mathbb{s} & 0 \\
 a \mathbb{c} & 0 & 0 & -a^2 \mathbb{c} \mathbb{s} ^2 \\
 -a r \mathbb{s} & 0 & 0 & r \mathbb{s} \left(a^2+r^2\right) \\
 0 & a^2 \mathbb{c} \mathbb{s} ^2 & -r \mathbb{s} \left(a^2 + r^2\right) & 0 \\
\end{array}
\right).
\end{equation}
%

For our purposes, these are combined into a third conformal Killing-Yano 2-form, $\hat Z := \hat H + \imath \hat F$,  with the corresponding conserved quantity the vector $\mathbb{k}$ of Eq. \ref{xi}:
\begin{equation}\label{xi}
{\bf f} \cdot \hat Z \cdot {\bf p} = \mathbb{k}.
\end{equation}

Conservation of the Fermi-Walker constant, $\mathbb{k}$, along the spacetime geodesics of Eq. \ref{traj1}, allows the polarization of light to be determined along any such path.  Since $\mathbb{k}$ is complex valued, its real and imaginary components provide two equations for the four components of the polarization vector, $\bf f$, provided that the initial polarization is known. To this end, Eq. \ref{k_Kerr} is decomposed into real and imaginary parts:
\begin{equation}\label{k_KerrRI}
{\mathbb k}_R({\bf f}, {\bf p}, \theta) = r \alpha -a \beta \mathbb{c} , \quad
{\mathbb k}_I({\bf f}, {\bf p}, \theta)  = -( r \beta + a \alpha  \mathbb{c} ).
\end{equation}
The terms $\alpha$ and $\beta$, defined in Eq. \ref{k_Kerr_2}, are functions of polar angle, $\theta$, polarization, $\bf f$, and the unit tangent to the trajectory, $\bf p$. Conservation of $\mathbb{k}$ then implies that
\begin{align}\label{eqs1and2}
{\mathbb k}_R({\bf f}_{\scriptscriptstyle \rm init}, {\bf p}_{\scriptscriptstyle \rm init}, \theta_{\scriptscriptstyle \rm init}) &= {\mathbb k}_R({\bf f}(s), {\bf p}(s), \theta(s) ) \nonumber \\
{\mathbb k}_I({\bf f}_{\scriptscriptstyle \rm init}, {\bf p}_{\scriptscriptstyle \rm init}, \theta_{\scriptscriptstyle \rm init})  &= {\mathbb k}_I({\bf f}(s), {\bf p}(s),  \theta(s) ).
\end{align}

A third equation is generated by the orthogonality of polarization and the wave vector---i.e. ${\bf f} \cdot {\bf p} = 0$: 
\begin{equation}\label{eq3}
f^3 = \frac{\Sigma^2 \left(\Delta  f^2 p^2 + f^1
   p^1\right)}{\mathbb{s}^2 \Delta \left(2
   a p^0 r - p^3
   \left(\left(a^2 + r^2\right)^2-a^2 \mathbb{s}^2 \Delta \right)\right)}.
\end{equation}

Finally, tangents to the trajectory are null vectors, so the polarization is only unique modulo a factor of the wave vector. The factor can always be chosen so that the temporal component of the polarization is equal to zero:
\begin{equation}\label{eq4}
f^{0}(s) = 0.
\end{equation}
Given a trajectory and initial polarization, $f^3$ and $f^0$ of Eqs. \ref{eq3} and \ref{eq4} can be substituted into Eqs. \ref{eqs1and2} which are subsequently solved for $f^1$ and $f^2$ as a function of Mino time, $s$, along the trajectory. This is a primary application of the Walker-Penrose theorem. The final pair of equations are linear in $f^1$ and $f^2$, but the analytical expressions obtained are unwieldy and lacking in immediate physical insight on their own. Since they are easily obtained with symbolic algebra software, the explicit expressions are not written here. They are used, though, in the application to follow.

\subsection{Gravitational Faraday Rotation}

A methodology has been laid out for analytically determining the trajectory of light and the evolution of its polarization. The latter can be measured at the start and end of a closed path with spacetime foliated into a curved three-space and a temporal path constructed so that its tangent is the temporal Killing vector, ${\bf v}_t$ of Eq. \ref{Killing_Vectors}. Projection of the trajectory and evolving polarization into three-space results in a path that is not a geodesic and a polarization that is not parallel transported. It is this projection that results in a dynamic in which linear polarization appears to be rotating due to the presence of an external influence, a Coriolis force that can be interpreted as arising from a fictitious magnetic field. 

To quantify what is perceived as an anomalous rotation, it is important to appreciate that the parallel transport of polarization in four-space exhibits the minimum possible vector rotation about the propagation direction that is consistent with all other constraints. Any twisting of the polarization is the result of path curvature. Its projection in three-space, though, appears to rotate in a way that does not exhibit minimal rotation. This can be quantified by constructing an orthonormal triad at each point along the path that does, indeed, evolve with minimal rotation. The rotation of the projected polarization vector, relative to this reference frame, is the gravitational Faraday rotation (GFR). Evolution of the reference triad is referred to as Fermi-Walker transport, which amounts to a set of ordinary differential equations very similar to those of parallel transport:
\begin{equation}\label{FWX}
\frac{D f^i}{D s} = \left( \frac{D n^i}{D s} n^k - \frac{D n^k}{D s} n^i \right) f^{\rm (proj)}_k .
\end{equation}
Here $f^i$ are the components of one of the reference triad vectors, $\bf f$, the three-space unit tangent to the trajectory is $\bf n$, 3-space vector $\bf f^{\rm (proj)}$ is the three-space projection of $\bf f$, and $D$ is the three-space covariant derivative. 

A three-space metric can then be constructed so that the covariant derivative can be evaluated. The three-space metric is then
\begin{equation}\label{gamma}
\gamma_{i,j} = \frac{g_{ij}}{h} + g_i g_j .
\end{equation}
Scalar field $h$ and vector field ${\bf g}$ are referred to as the \emph{gravitational electrostatic potential} and \emph{gravitational magnetic vector potential}, respectively\cite{Fayos_1982}. These names are intended to identify the fields as analogs to counterparts in Maxwell theory relevant to Faraday rotation.  Three-vector ${\bf g}$ has only a $\phi$-component, so its curl (analogous to an inhomogeneous, static, magnetic field, ${\bf B}$) is pointed along the rotational axis of the black hole. Faraday rotation therefore occurs when linearly polarized radiation has a propagation unit vector, ${\bf n}$, that has a non-zero axial component. The vector ${\bf n}\times{\bf g}$ can be interpreted as a Coriolis force. 

If the only objective is to quantify GFR, then it is possible to sidestep the need to solve the Fermi-Walker transport equations, instead drawing on an elegant analysis by Frolov and Shoom\cite{FrolovShoom2011}. They show that GFR can be expressed as the integral of a rotation rate, $\Omega$, given by
\begin{equation}\label{Frolov}
\GFR(s) = \int_0^{s} d{\mathbb s}\, \Omega_{\rm GF}({\mathbb s}),
\end{equation}
where
\begin{equation}\label{Omega}
\Omega_{\rm GF}(s) = \frac{1}{2} {\rm curl}{\bf g}(s) \cdot {\bf n}(s). 
\end{equation}
However, this does not allow us to visualize the Fermi-Walker transport of polarization and develop an intuitive sense of how to engineer GFH. It will therefore be used only as a check on our numerical solution to the Fermi-Walker evolution of polarization.

\section{Application}

\subsection{Trajectories}

\begin{figure}[t]
	\begin{center}
		\includegraphics[width=1.00\linewidth]{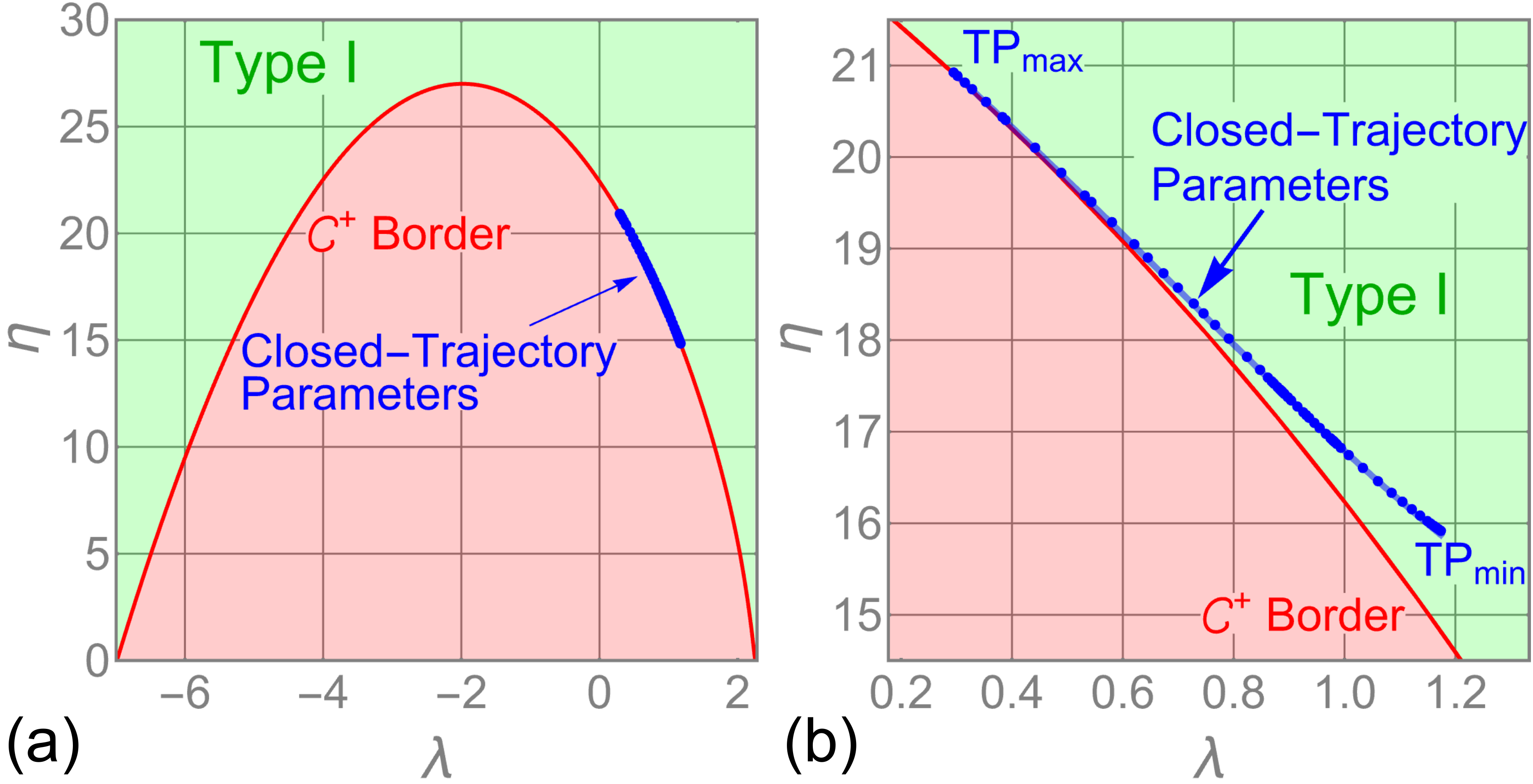}
	\end{center}
	\caption{ \emph{$\cal{C}_+$ Boundary and $\{\lambda,\eta\}$ pairs.} The $\cal{C}_+$ border, shown in dark red, is associated with orbits of constant Boyer-Lindquist radius. Killing ratios $\lambda$ and $\eta$ are chosen to be in the region above the $\cal{C}_+$ border.  These are shown in blue points in both panels. Position within the Type I region (green) is scrutinized in panel (b), where only the relevant domain is plotted and the vertical difference between each blue data point and the red $\cal{C}_+$ curve has been multiplied by a factor of 50. This makes it easier to see that the blue points approach $\cal{C}_+$ as $\lambda$ decreases. The lighter blue curve behind these points is a fit for which the vertical distance up from the $\Cplus$ border is $\lambda^4/90$. Labels ${\rm TP}_{\rm min}$ and ${\rm TP}_{\rm max}$ identify the parameters, associated with the minimum and maximum turning point radii, for which a closed trajectory is possible. $a = 0.99.$} 
	\label{Cplus}
\end{figure}

While the equations of motion must be solved to produce specific optical trajectories, their character can be anticipated by examining where the associated impact parameter and Carter ratio lie on the parameter plane. For instance, parameter pairs that lie on the $\Cplus$ border, shown in red Fig. \ref{Cplus}(a), produce trajectories that have a constant Boyer-Lindquist radius\cite{Lusk_PRD_2024}. Our focus, though, is within the green "Type I" region that lies above this curve. It is defined by the character of the roots of the radial potential, ${\cal R}$, given in Eq. \ref{Rpotential}. This amounts to a quartic polynomial in radial position, $r$, for which choices of $\lambda$ and $\eta$ affect the values of its four solutions.  In the Type I region, all four roots are real, with two inside the inner event horizon and two outside the outer event horizon. If the source/receiver radial position, $\rsr$, is greater than the largest of these, then the trajectory will have a single radial turning point, $\rTP$, at that largest root. The polar angle may have many turning points while the azimuthal angle will have none, and trajectories will exhibit nontrivial scattering.

Within this Type I region, we seek a specific subtype of trajectories for which the path self-intersects. Identification of the requisite parameters was guided by the following considerations:

\begin{enumerate}[labelwidth=!, labelindent=0pt, leftmargin=10pt]

\item{\emph{Paths exterior to ergosphere.} A (3+1) foliation of spacetime is necessary for the  construction of an evolving Fermi-Walker frame of reference.} 

\item{\emph{Axial motion.} GFR is only generated along paths that have a component parallel to the z-axis, generally associated with a large Carter constant, $Q$. Recall that $\eta = Q/\varepsilon$, so $\eta$ should be large.}

\item{\emph{Parameters near $\Cplus$.} $\rTP$ increases away from the $\Cplus$ border, but it should be small so that light traverses spacetime regions that are substantially curved. Otherwise the trajectory will not close.}

\item{\emph{Trajectory asymmetry.} The rises and falls of GFR are unlikely to cancel if the path is asymmetric.}

\end{enumerate}

The source/receiver position was chosen to lie within the equatorial plane with $\phi = 0$, so that its location is completely determined by the radial position, $\rsr$. Guided by the considerations above, an initial parameter pair, $\{\lambda,\eta\}$, was identified by trial and error to produce a closed trajectory for which $\phi(\smax) = 3*2\pi$. This is the middle (tan) curve shown in Fig. \ref{Trajectories}, for which $\rsr = 4.57$. The source radius was then increased/decreased in small increments, with a root finder used to determine new values of $\{\lambda, \eta, \smax\}$ such that $r(\smax) = r(0)$, $\theta(\smax)= \pi/2$, and $\phi(\smax) = 3*2\pi$. 

Fig. \ref{Cplus}(b) shows the resulting set of 65 data points (blue). It is bounded at left where it touches the $\Cplus$ curve, and at the right where the turning point radius is equal to the radius of the ergosphere at the equator. The set of points is fitted reasonably well by 
\begin{equation}\label{fit}
\Delta\eta = \lambda^4/90,
\end{equation}
with $\Delta \eta$ the vertical distance between the $\Cplus$ curve and each blue data point. This fit is shown in light-blue in Fig. \ref{Cplus}.

%
%
%

\subsubsection{Closed Trajectories}

\begin{figure}[t]
	\begin{center}
		\includegraphics[width=0.7\linewidth]{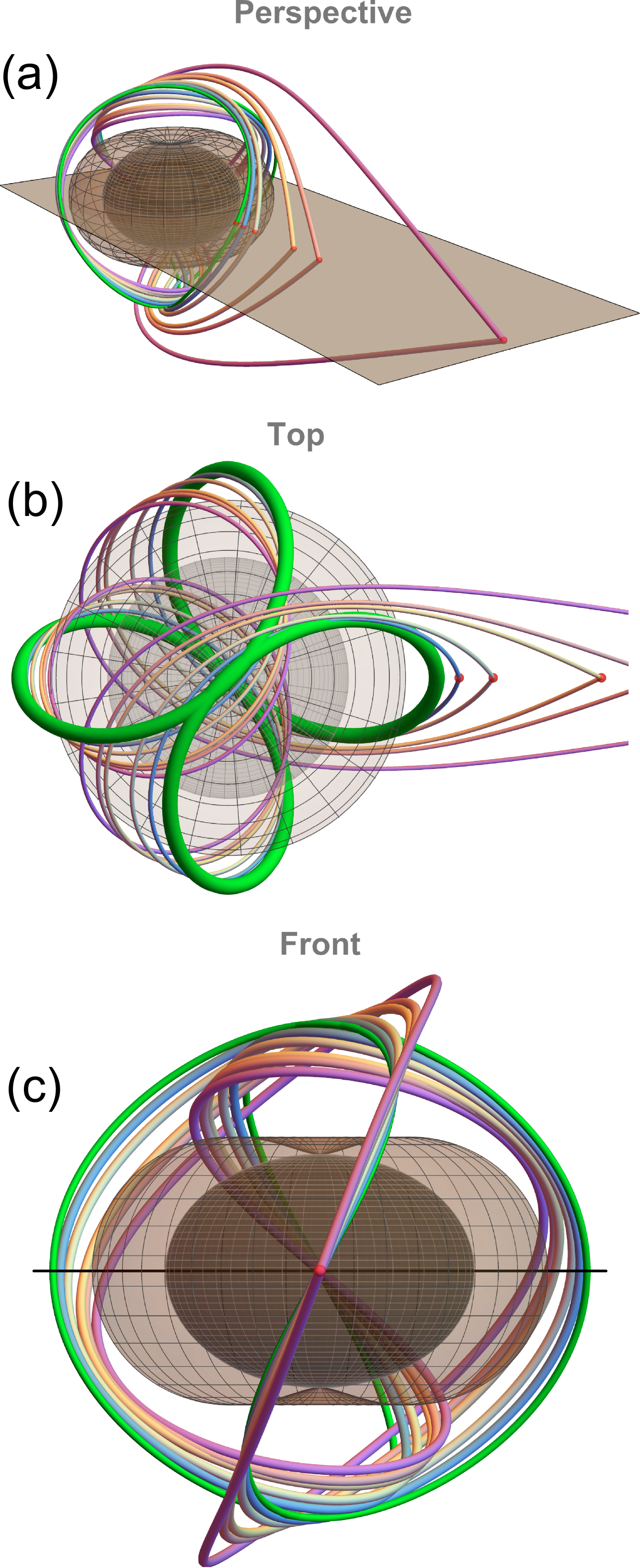}
	\end{center}
	\caption{ \emph{Trajectories.} Representative three-space trajectories are plotted for $\rsr = 3.17, 3.42, 4.57, 5.43$, and $10.5$. The trajectory shown in green is iso-radial, with Boyer-Lindquist radius = 2.42. Its parameters lie on the red $\cal{C}_+$ curve of Fig. \ref{Cplus} with $\lambda = 0.0334, \eta = 22.3$.  The outer event horizon is shown in dark gray while the outer ergosphere surface is depicted with a lighter gray. $a = 0.99$. Perspective, top, and front views are shown in panels (a), (b), and (c), respectively.}
	\label{Trajectories}
\end{figure}

A closed trajectory was generated for each of the blue data points shown in Fig. \ref{Cplus}. A representative selection of these is shown in Fig. \ref{Trajectories}. Also plotted there (thick green) is a closed circuit for which conserved parameters $\lambda$ and $\eta$ lie on $\Cplus$. This produces a 4-petal closed trajectory on a surface of constant Boyer-Lindquist radius. Its symmetry is apparent in Fig. \ref{Trajectories}(b), and it is this that results in a dearth of holonomy. As the source/receiver radius is increased, though, this symmetry is lost in both the individual petal shapes and their three-space orientation. The degree of asymmetry increases as the source is moved outward.

%
%

\begin{figure}[t]
	\begin{center}
		\includegraphics[width=1.00\linewidth]{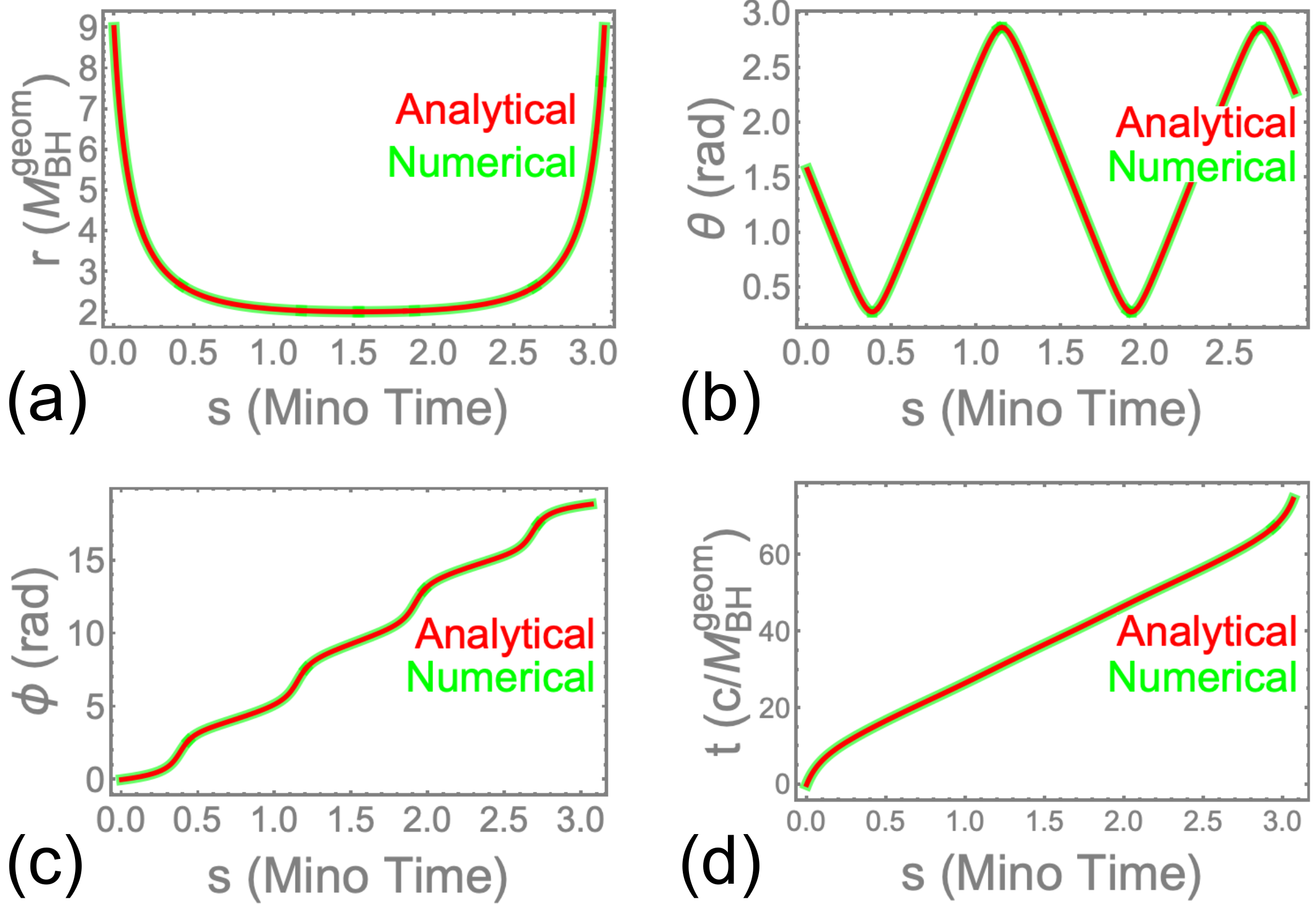}
	\end{center}
	\caption{ \emph{Comparison of Analytical and Numerical Geodesics}. Analytical solutions to the geodesic equations of motion---i.e. PX of the wave vector---are compared with counterparts obtained numerically. The plots exhibit a high degree of self-consistency, with an average discrepancy on the order of $10^{-4}\%$. Here $a = 0.99$, $\rsr = 9.0$.} 
	\label{NumericalChecks}
\end{figure}

As a check on the analytical expressions of Eqs. \ref{traj1}, equations of motion \ref{EoMr}--\ref{EoMt} were also solved numerically in each case. An example is given in Fig. \ref{NumericalChecks}, where discrepancies are on the order of $10^{-4}\%$, a level that can be attributed to the numerical solver. This level of validation is consistent with all 65 simulations corresponding to the blue points in Fig. \ref{Cplus}. It was the initial comparison effort that uncovered several issues with the expressions for $\phi(s)$ and $t(s)$ provided by Gralla \cite{Gralla_2020}. These were subsequently corrected, with the validated expressions provided in the Appendix A.

%
%
%

\subsection{Parallel Transport of Planes of Polarization}

With a set of trajectories now constructed, attention is turned to the evolution of a pair of orthogonal, linear polarizations, ${\bf A}(s)$ and ${\bf B}(s)$. Their initial state is determined by constructing a real-valued, orthogonal tetrad using wave vector, ${\bf p}$. First identify the temporal and longitudinal polarization vectors, 
\begin{equation}
{\bf T} = \frac{i {\bf v}_t}{ |{\bf v}_t| } , \quad
{\bf L} = \frac{{\bf p} + {\bf T} ({\bf p}\cdot {\bf T})}{ |{\bf p}\cdot {\bf T}|} ,
\end{equation}
with ${\bf v }_t$ the temporal Killing vector of Eq. \ref{Killing_Vectors}. Their lengths are $-1$ and $+1$, respectively. Pairwise orthogonality of the tetrad vectors then gives five scalar equations for the unknown components of ${\bf A}(0)$ and ${\bf B}(0)$. Two additional scalar equations are obtained by insisting that they be of unit length. A final equation is provided by requiring that the polar component of ${\bf B}(0)$ be equal to zero. This is arbitrary but useful. The algebraic equations are then solved to obtain ${\bf A}(0)$ and ${\bf B}(0)$. 

Finally, tangents to the trajectory are null vectors, so these polarizations are only unique modulo a factor of the wave vector. As noted in Eq. \ref{eq4}, the factor can always be chosen so that the temporal component of ${\bf A}(0)$ and ${\bf B}(0)$ are equal to zero:
\begin{align}
{\bf A}_{\rm phys}(0) &= {\bf A}(0) - \frac{A^0(0)}{p^0(0)} {\bf p}(0) \nonumber \\
{\bf B}_{\rm phys}(0) &= {\bf B}(0) - \frac{B^0(0)}{p^0(0)} {\bf p}(0) .
\end{align}
These are the physically meaningful polarization vectors to be parallel transported.  

As will be verified subsequently, GFR is agnostic with respect to polarization, so polarization $\bf A$ is chosen for the sake of definiteness. The orthogonality of generic polarization $\bf A$ and wave vector $\bf p$ implies that the polarization is parallel transported:
\begin{equation}\label{PX}
({\bf u} \cdot\nabla){\bf A} \equiv \nabla_{\bf u}{\bf A} = 0 .
\end{equation}
This is equivalent to the following indicial equalities:
\begin{equation}\label{PX}
\frac{D A^\mu}{D s} \equiv \frac{DA^\mu}{D x^\nu}\frac{d x^\nu}{ds} = 0, 
\end{equation}
where $D$ is the four-space covariant derivative. As verified numerically, the solution to these equations is equivalent to that obtained using the Walker-Penrose theorem---i.e. by applying Eqs. \ref{eqs1and2}, \ref{eq3}, and \ref{eq4}.

Once the polarization vector has been parallel transported, it is projected into three-space, $\APX$, by removing its temporal component. This component is initially zero but will be finite along the trajectory in general. An evolving plane of polarization, $\PPX$, can then be constructed by taking the cross product of the unit tangent to the three-space trajectory, $\bf n$, and $\APX$:
\begin{equation}\label{PoP}
P_A^{\!\scriptstyle\rm (PX3),i}:= \frac{1}{\sqrt{{\rm det}(\gamma)}} \epsilon^{ijk} n_j A^{\!\scriptstyle\rm (PX3)}_k . 
\end{equation}
Here $\epsilon^{ijk}$ is the Levi-Civita symbol and $\gamma$ is the three-space metric given in Eq. \ref{gamma}. The angle between the parallel transported (PX) and Fermi-Walker transported (FWX) plane of polarization vectors is due to GFR. It is important to note that projection and transport operations do not commute. 

\begin{figure}[t]
	\begin{center}
		\includegraphics[width=1.00\linewidth]{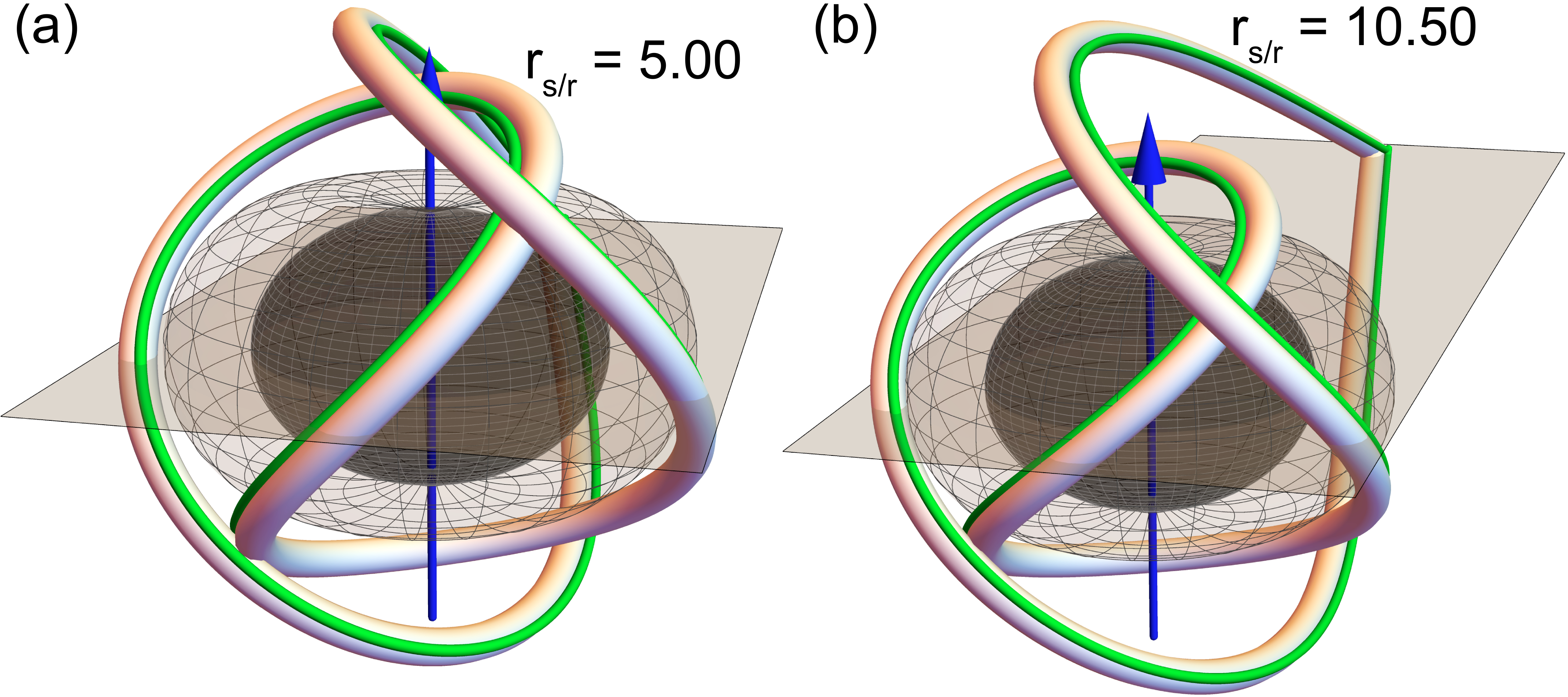}
	\end{center}
	\caption{ \emph{Parallel Transport of Planes of Polarization}. Parallel transport of three-space polarization is shown for two choices of source/receiver radius, $\rsr$. Trajectories are portrayed as thick ivory paths, with polarization represented as a smooth green curve that connects the tips of vector along a given path. The qualitative character of the polarization curves does not change with source/receiver position. $a = 0.99$.} 
	\label{PX_PoP}
\end{figure}

The PX-evolving plane of polarization is shown in Fig. \ref{PX_PoP} for two values of source radius. While there are differences in the way in which the plane evolves, these are relatively subtle and do not exhibit qualitative differences. The same cannot be said when the vector is FWX, and that is taken up next.

%
%
%

\subsection{Fermi-Walker Transport of Planes of Polarization}

The three-space plane of polarization for ${\bf A}(s)$ can be FWX by numerically solving Eq. \ref{FWX}. The initial state is given by $\APX(0)$. The equations are similar to those of FWX, Eq. \ref{PX}, but the FW equations are in three-space instead of four-space. While the equations offer additional insight when expressed in terms of a specially chosen tetrad frame\cite{Tanaka_1996}, they would still need to be solved numerically. That additional step was therefore not adopted. 

\begin{figure}[t]
	\begin{center}
		\includegraphics[width=1.00\linewidth]{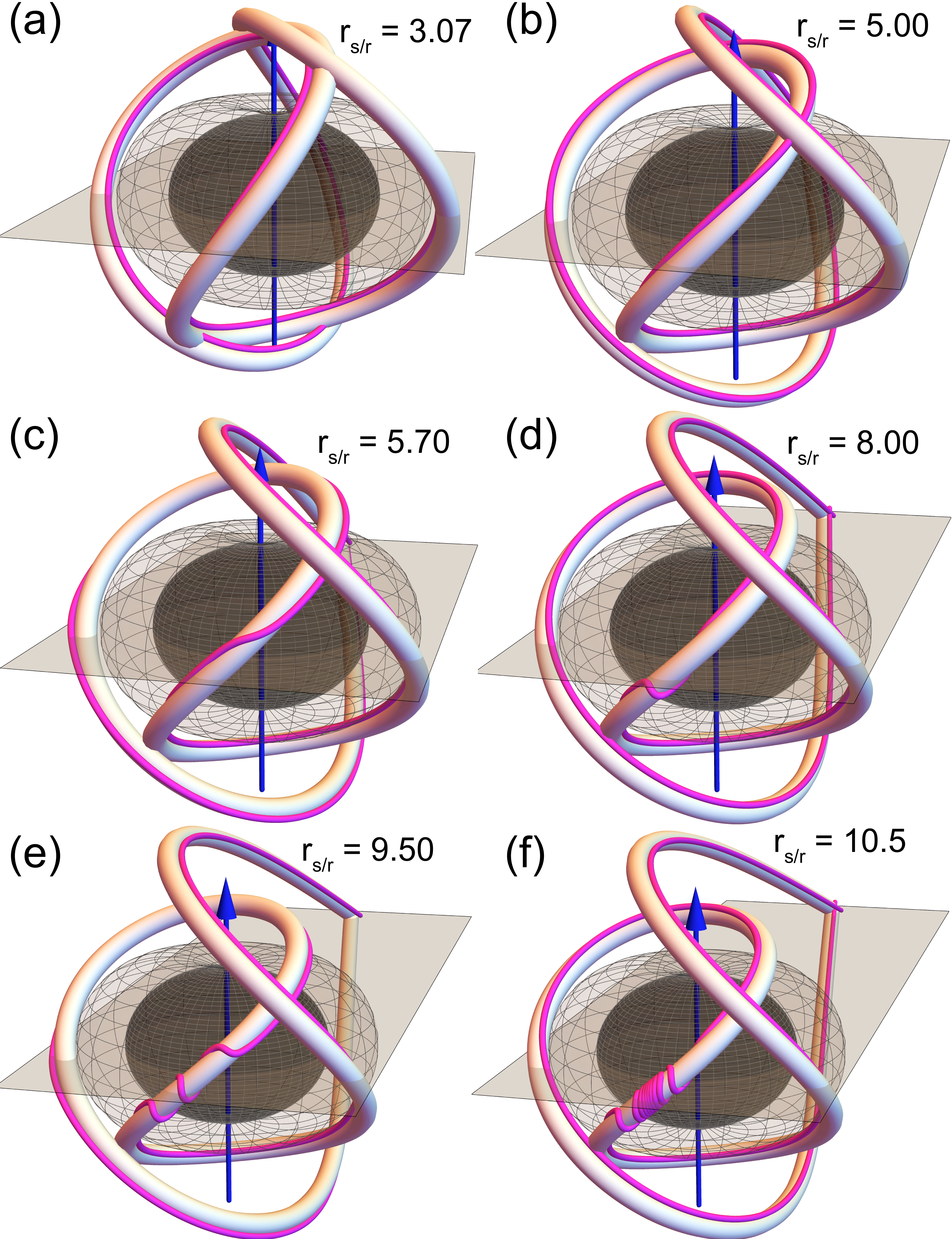}
	\end{center}
	\caption{ \emph{Evolution of Polarization}. Fermi-Walker transport of three-space polarization is shown for several choices of source/receiver radius, $\rsr$. Trajectories are portrayed as thick ivory paths, with polarization represented as a smooth magenta curve that connects the tips of vector along a given path. All trajectories exhibit a radial turning point in the equatorial plane (foreground), and the FW polarization is observed to spiral in this region. The degree of spiraling increases as the source/receiver is moved out radially. $a = 0.99$.} 
	\label{FW_PoP}
\end{figure}

The FWX plane of polarization is plotted in Fig. \ref{FW_PoP} for six values of source radius. This helps to see the qualitative changes that develop as the source is moved further outwards. At relatively small source radii, the FW trajectories look very similar to their PX counterparts, as in Fig. \ref{PX_PoP}. As the source is moved further away from the singularity, though, the vector exhibits an increasingly severe spiral about the optical path. This occurs in a region for which the light makes an equatorial transit and for which the sole radial turning point is present. In fact, the ambits were engineered to exhibit just such a behavior, a consequence of the design rules given earlier. This spiraling has significant implications for GFH, as will be discussed next. 

%
%
%

\subsection{Gravitational Faraday Rotation and Holonomy}

The underlying physics can now be explained with reference to Figs. \ref{PX_PoP} and \ref{FW_PoP}. Light follows a geodesic path in four-space and its polarization is parallel transported there. In that setting, parallel transport is equivalent to Fermi-Walker transport---i.e. the polarization vector rotates about the propagation axis to the minimum possible. Projection of the trajectory and evolving polarization into three-space results in a path that is not a geodesic and a polarization that is not parallel transported. It is this projection that results in a dynamic in which its plane of polarization, $\PPX(s)$, appears to be rotating due to the presence of an external influence, a Coriolis force. This perceived rotation can be measured by comparing the three-space plane of polarization with its FW-evolving counterpart, $\PFW(s)$:
\begin{equation}
\GFR(s) = \cos^{-1}\left(\PPX(s) \cdot \PFW(s) \right) .
\end{equation}
This gives a result identical to that of Eq. \ref{Omega}. The motivation for our numerical procedure is that the planes of polarization can be visualized, useful in both understanding GFR accumulation and in engineering ambits for which a significant gravitational Faraday holonomy can be realized. 

It is worth noting that the single discontinuity in the tangent vector, at the source/receiver, is irrelevant to GFH. This is not surprising as it is consistent with more standard measurements of geometric phase. For instance, the parallel transport of a vector tangent to the surface of a sphere results in a misorientation between start and end configurations equal to the negative of the solid angle enclosed, even when the tangent vector has one or more discontinuities.

\begin{figure}[t]
	\begin{center}
		\includegraphics[width=0.8\linewidth]{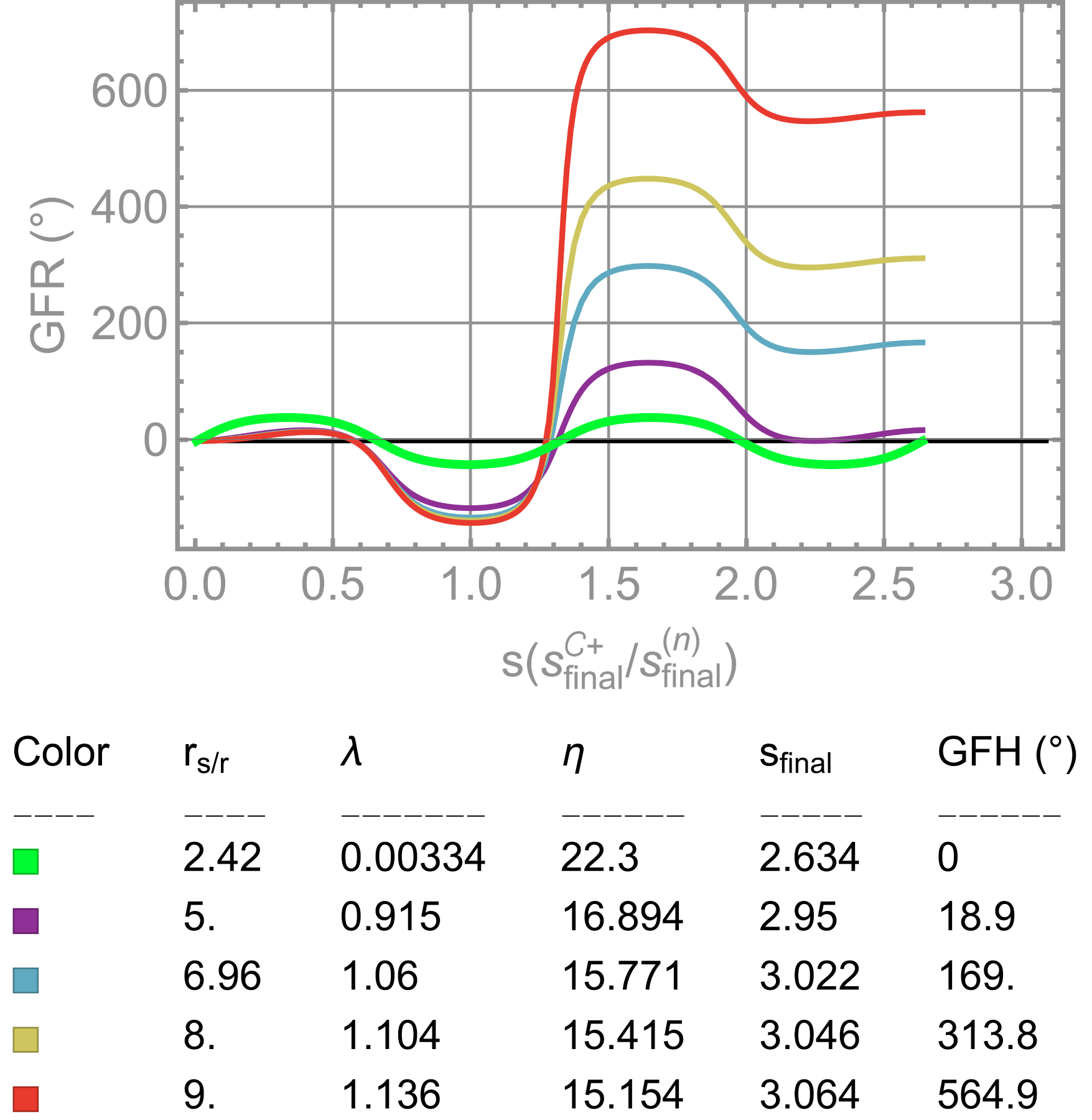}
	\end{center}
	\caption{ \emph{GFR Accumulation}. Gravitational Faraday Rotation is plotted, as a function of path position, for a range of relatively small source/receiver positions, $\rsr$. The green curve, associated with the spherical trajectory shown in green in Fig. \ref{Trajectories}. $a = 0.99$.} 
	\label{GFR_v_s_Comparison}
\end{figure}

The accumulation of GFR is shown, in Fig. \ref{GFR_v_s_Comparison}, for several source positions. First focus on the green curve, associated with the spherical trajectory, also in green, of Fig. \ref{Trajectories}. It exhibits regions of increasing and decreasing GFR which ultimately cancel, consistent with the 4-petal path symmetry described earlier. Trajectories associated with non-spherical orbits do not have such symmetry, though, and Fig. \ref{GFR_v_s_Comparison} shows that this is manifested in a sharp rise in GFR at the midpoint that increases with $\rsr$. The result is a gravitational Faraday holonomy (GFH). 

%
%
%

\begin{figure}[t]
	\begin{center}
		\includegraphics[width=0.8\linewidth]{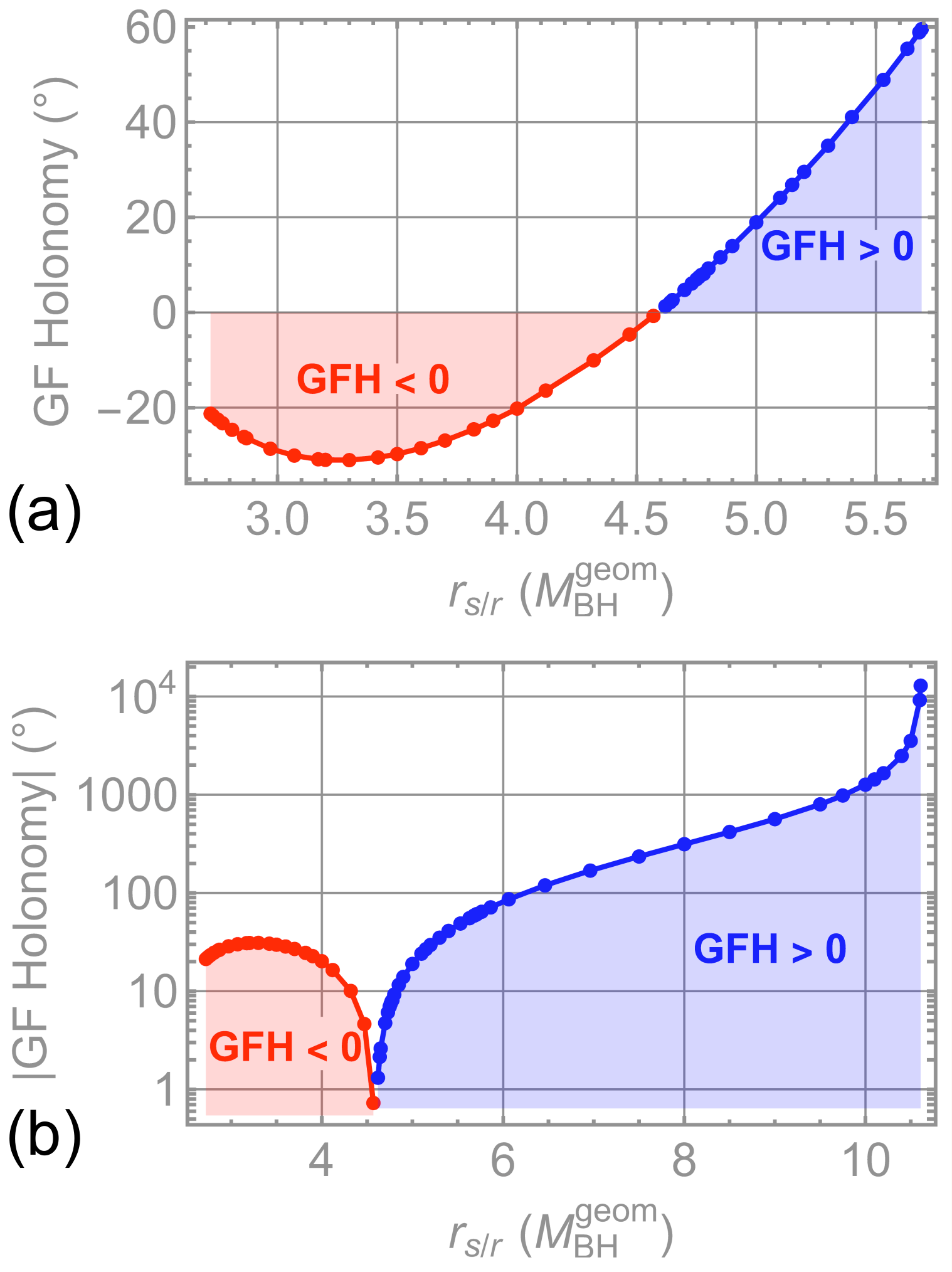}
	\end{center}
	\caption{ \emph{GFH Versus Source/Receiver Radius}. (a) Gravitational Faraday Holonomy (GFH) is plotted over a range of relatively small source/receiver positions, $\rsr$. (b) The range of $\rsr$ is extended and, because the holonomy becomes so large, the absolute value is plotted on a log scale to show how it grows with radial position of the source/receiver. $a = 0.99$.} 
	\label{GFH_v_Radius}
\end{figure}

The entire spectrum of closed trajectories is used to quantify GFH as a function of source position. Fig. \ref {GFH_v_Radius} indicates that it is negative when the source is relatively close, before turning positive for more distant sources. The holonomy eventually undergoes an excursion, growing so large that it is plotted on a log scale in panel (b).  While it is reasonable to posit that this is related to the polarization spiraling of Fig. \ref{FW_PoP}, the root cause bears investigation.

%
%
%

\section{Explanation for GFH Excursion}

\begin{figure}[t]
	\begin{center}
		\includegraphics[width=0.9\linewidth]{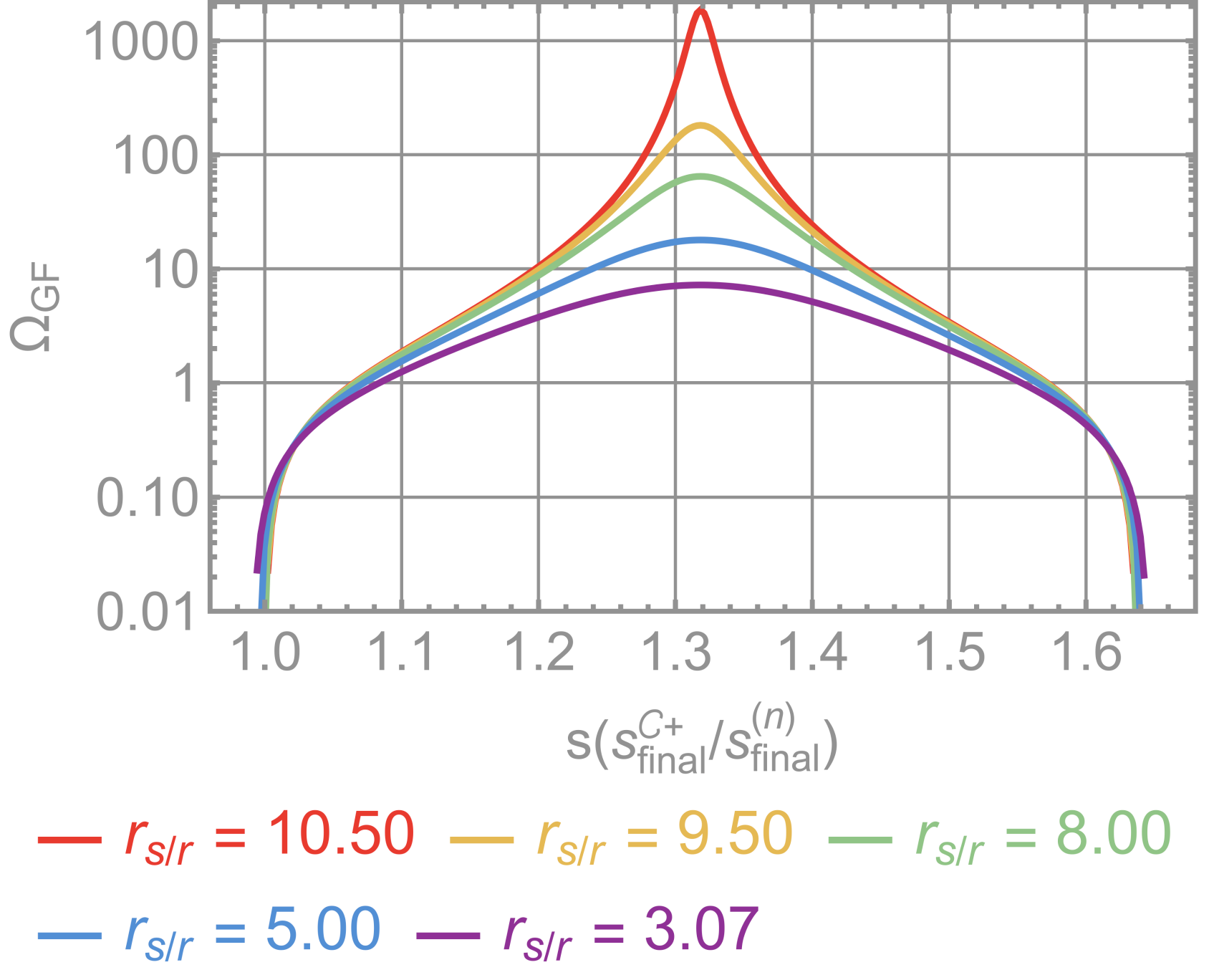}
	\end{center}
	\caption{ \emph{GF Rotation Rate}. The rate of change of GFR with respect to Mino time is plotted for representative positions of the source/receiver. $a = 0.99$.} 
	\label{GFR_Rates}
\end{figure}

As a first step in elucidating the cause of the holonomy excursion, Faraday rotation rate, $\Omega_{\rm GF}$, of Eq. \ref{Omega}, is plotted as a function of Mino time in Fig. \ref{GFR_Rates}.  The log scaling makes it clear that this rate is only significant at the mid-trajectory segment that passes through the equatorial plane, which becomes increasingly severe as the source/receiver is moved outwards.

%
%
%

\begin{figure}[t]
	\begin{center}
		\includegraphics[width=0.7\linewidth]{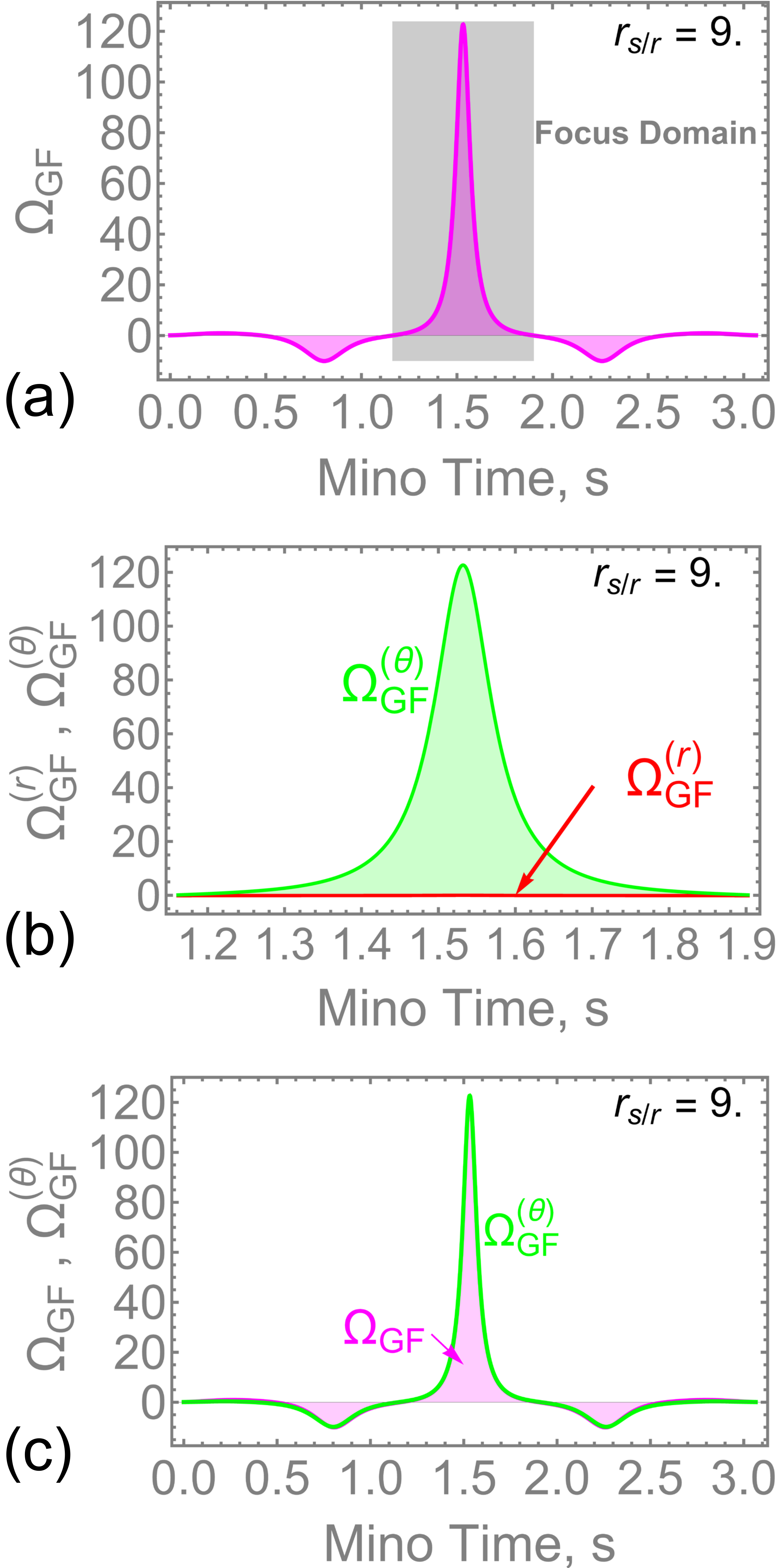}
	\end{center}
	\caption{ \emph{$\Omega_{\rm GFR} \approx  \Omega_{\rm GFR}^{(\theta)}$}.  (a) The GFR accumulation, plotted in Fig. 
\ref{GFR_v_s_Comparison}, was obtained by integrating the Faraday rotation rate, $\Omega_{\rm GF}$ over the trajectory. This rate is plotted to make clear that it is only significant over the mid-trajectory segment that passes through the equatorial plane. (b) The rotation rate can be decomposed into two contributions, given in Eq. \ref{2contribs}. Only the polar contribution (green) is significant over the domain of interest. (c) The total rotation rate is well-approximated, over the entire trajectory, by its polar contribution. $a = 0.99$.} 
	\label{Explanation_1}
\end{figure}

With this in mind, consider a specific trajectory ($\rsr = 9.0$) to deconstruct the rotation rate.  Consistent with the rapid rise in GFR at the circuit midpoint, Fig. \ref{Explanation_1}(a) shows the Faraday rotation rate, $\Omega_{\rm GF}$, on a linear scale to fully appreciate how rapidly it rises. It can be expanded out as
\begin{equation}
\Omega_{\rm GF} = \frac{1}{2} n_j (\nabla\times{\bf g})^j .
\end{equation}
Since $(\nabla\times{\bf g})^\phi = 0$ and the metric is diagonal in the Boyer-Lindquist chart, the rotation rate can be expressed as the sum of two contributions:
\begin{eqnarray}\label{2contribs}
\Omega_{\rm GF} &=&  \frac{1}{2} \gamma_{rr} n^r (\nabla\times{\bf g})^r + \frac{1}{2} \gamma_{\theta\theta} n^\theta (\nabla\times\bf{g})^\theta \nonumber \\
&\equiv& \Omega_{\rm GF}^{(r)} + \Omega_{\rm GF}^{(\theta)} .
\end{eqnarray}
As shown in Fig. \ref{Explanation_1}(b), though, the first term is negligibly small over the domain of interest.  
%
%
%
%
Therefore
\begin{equation}\label{Omega_approx}
\Omega_{\rm GF} \approx \Omega_{\rm GF}^{(\theta)} = \frac{1}{2} \gamma_{\theta\theta} n^\theta (\nabla\times\bf{g})^\theta .
\end{equation}
This is confirmed in Fig. \ref{Explanation_1}(c).

\begin{figure}[t]
	\begin{center}
		\includegraphics[width=0.7\linewidth]{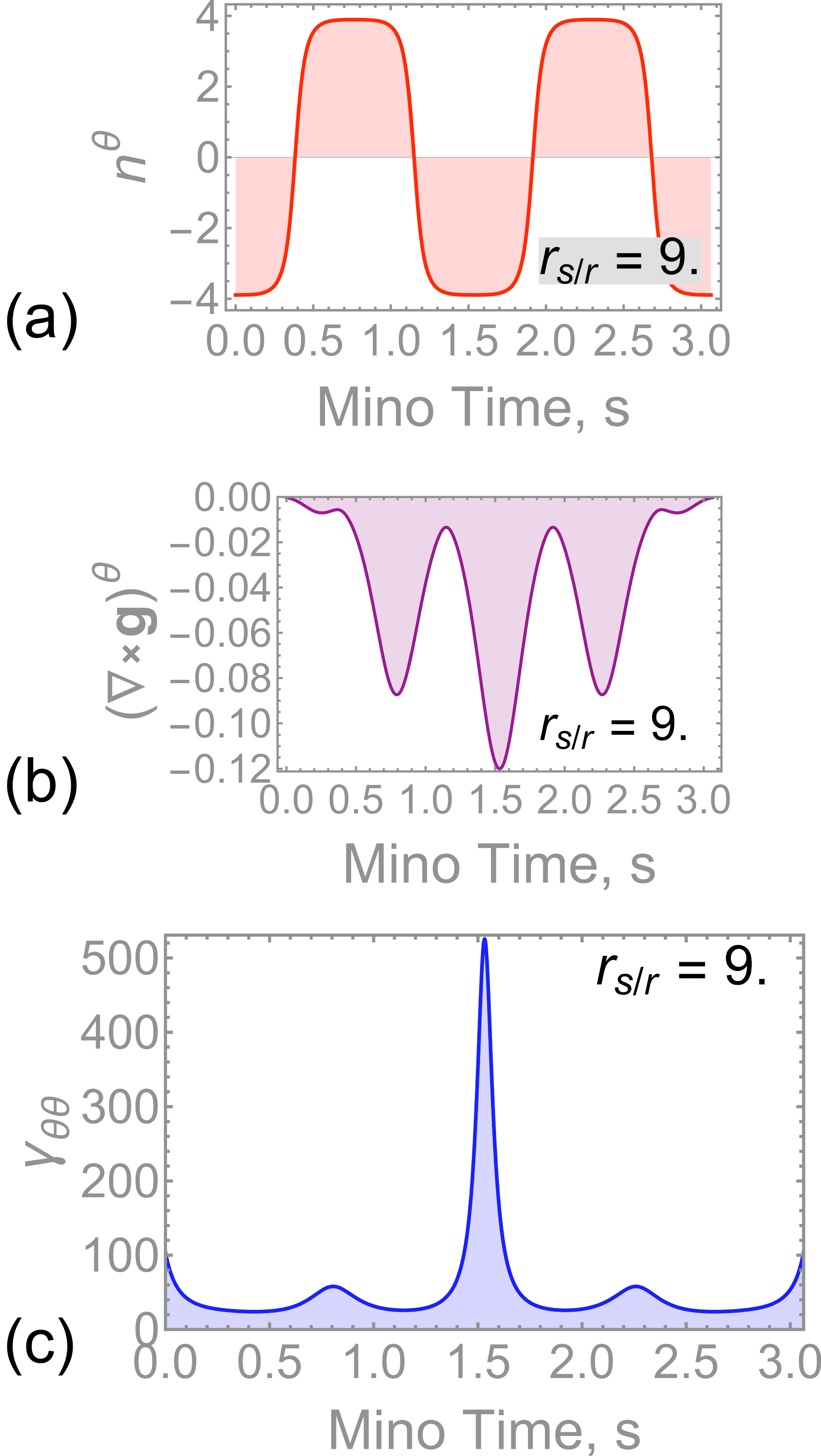}
	\end{center}
	\caption{ \emph{Excursion of $\Omega_{\rm GFR}$ and the three-space Metric.} As given in Eq. \ref{Omega_approx}, the rotation rate is the product of three terms: $n^\theta$, $(\nabla\times\bf{g})^\theta$, and  $\gamma_{\theta\theta}$. These are each plotted to determine that only $\gamma_{\theta\theta}$ exhibits an excursion as the trajectory transits its radial turning point at the equatorial plane. $a = 0.99$.} 
	\label{Explanation_2}
\end{figure}

The excursion in rotation rate, shown in Fig. \ref{Explanation_1}(b), must derive from one or more of the three terms at right in Eq. \ref{Omega_approx}. Two of these, $n^\theta$ and $(\nabla\times\bf{g})^\theta$, can be immediately ruled out by plotting them, and the source of the excursion as $\gamma_{\theta\theta}$ is confirmed in the same way. These are provided in Fig. \ref{Explanation_2}. 
%

%
%
%
%

\begin{figure}[t]
	\begin{center}
		\includegraphics[width=0.7\linewidth]{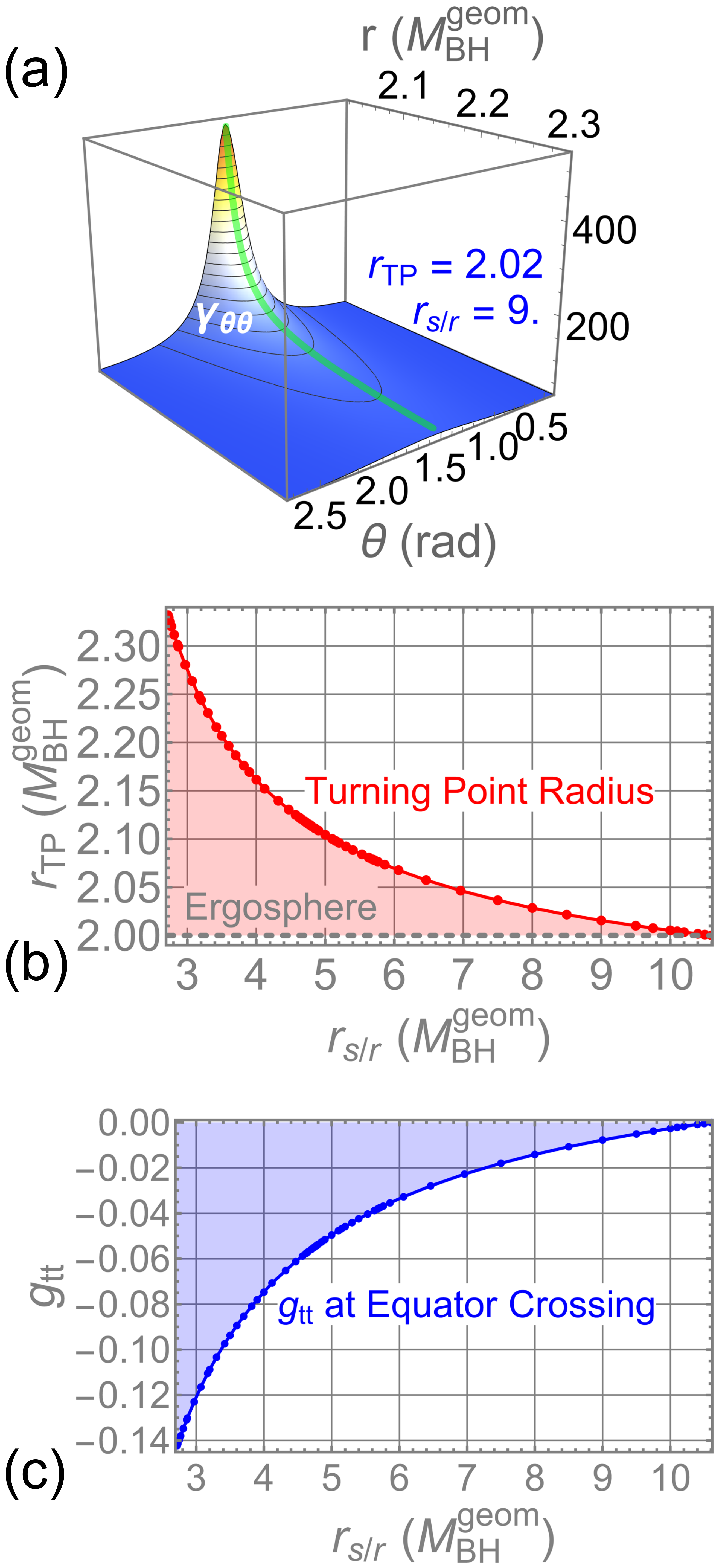}
	\end{center}
	\caption{ \emph{Collapse of Temporal Metric in Spacetime}. (a) The diagonal, polar component of the three-space metric is plotted as a function of polar angle and radial position for $a = 0.99$. It exhibits a peak at the equator and turning point radius. (b) The turning point radius is a function of the source/receiver position, and it approaches the equatorial transit becomes increasingly close to the ergosphere as the source/receiver is moved outwards. (c) Consistent with the discussion of Eq. \ref{gtt}, the $g_{tt}$ component of the BL metric approaches zero as the source/receiver radius is increased.} 
	\label{Explanation_3}
\end{figure}

The Kerr metric in the Boyer-Lindquist chart gives $\gamma_{\theta\theta}$ as
\begin{equation}\label{gamma_theta_theta}
\gamma_{\theta\theta} = \frac{ \left(r^2 + a^2 \cos^2\theta\right)^2 }{r(r-2) + a^2 \cos^2\theta}.
\end{equation}
This is plotted in Fig. \ref{Explanation_3}(a). The plot above shows that $\gamma_{\theta\theta}$ is large in only a small region for which $\theta$ is near the equator and $r$ is close to the trajectory turning point, $\rTP$. In this region, the metric component can be quite large, and it is this that is driving the rotation rate excursion. 

We have been careful to focus on a particular source/receiver radius, for the sake of clarity, but further insight requires that we consider trends associated with a range of source/receiver radii. In particular, we need to examine how the turning point itself varies, and this is given in Fig. \ref{Explanation_3}(b). As $\rsr$ increases, the turning point radius decreases, eventually encountering the ergosphere as it crosses the equatorial plane. This trend causes the peak in $\gamma_{\theta\theta}$ to be amplified, as plotted in Fig. \ref{Explanation_3}(a), and it becomes infinite when the ergosphere is reached.

The final step, an explanation for why $\gamma_{\theta\theta}$ increases as the ergosphere is approached, is inherent in its construction from the spacetime metric, $g$:
\begin{equation}\label{gtt}
\gamma_{\theta\theta} = \frac{1}{g_{tt}^2} \left( g_{t\theta}^2  - g_{\theta\theta}g_{tt} \right).
\end{equation}
As the ergosphere is approached, time dilation becomes increasingly severe, and $g_{tt}$ approaches zero. The three-space foliation is inversely proportional to this term and so becomes exceedingly compressed. 

It is worth noting that, for small source/receiver radii in the range of $\rsr < 3.3$, the GFH becomes increasingly negative as the source is moved outwards. This is because the initial distortion of the iso-radial 4-petal path, shown in green in Fig. \ref{Trajectories}(b), is more significant along the lower petal (top view), shifting clockwise but also moving closer to the ergosphere. The closest point of approach on this petal is always greater than the turning point radius, of course, and the accumulation of a negative contribution to GFH is eventually overwhelmed by the distortion at the turning point itself, which eventually grazes the ergosphere. 

%
%
%

\section{Physical Considerations for Holonomic Trajectories}

\subsection{Trajectory Length, Elapsed Time, and Time Dilation}

The trajectories we have been considering are closed because they pass through a region of spacetime close to the ergosphere. It is therefore reasonable to ask what sort of dilations are produced on such a path. In the Kerr metric, gravitational time and space dilations are distinct phenomena, each a function of position along the ambit of light. These can be quantified by expressing the space time interval, $dL$, in terms of Mino time derivatives:
\begin{align}\label{dL3}
\left(\frac{dL}{ds}\right)^2 &= \frac{\Sigma}{\Delta}\left( \frac{dr}{ds} \right)^2  + \Sigma\left( \frac{d\theta}{ds} \right)^2  \nonumber \\
& \!\!\!\!\!\!\!\!\!\!+ \frac{\sin^2\theta \left( (a^2 + r^2)^2 - a^2 \Delta\sin^2\theta \right)}{\Sigma}\left( \frac{d\phi}{ds} \right)^2  .
\end{align}
 Take the positive square root and integrate to quantify the total path length from the source to an arbitrary point, $s$, on the trajectory:
\begin{equation}\label{L3}
L(s) = \int_0^s ds \frac{dL}{ds} .
\end{equation}

\begin{figure}[t]
	\begin{center}
		\includegraphics[width=0.8\linewidth]{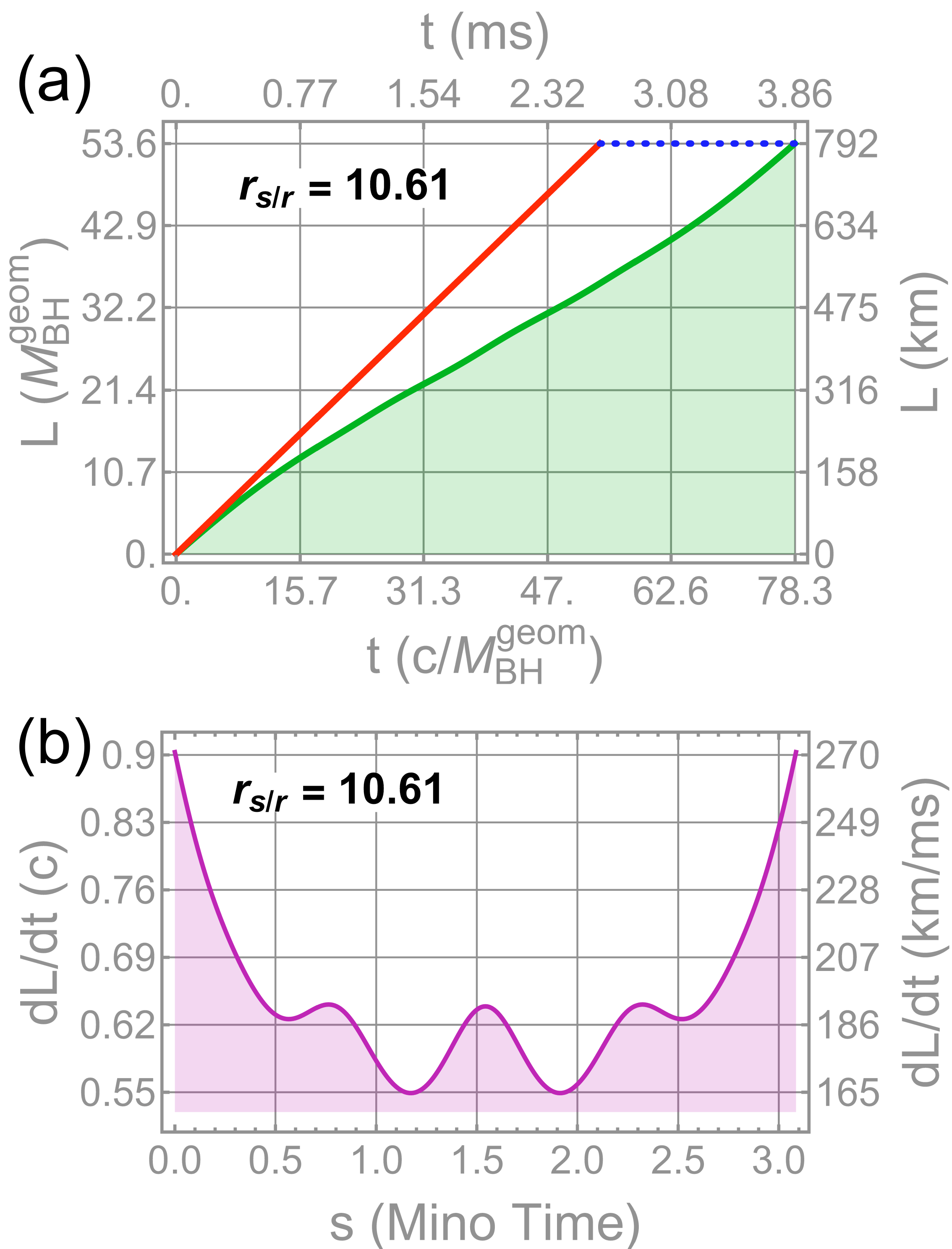}
	\end{center}
	\caption{ \emph{Total Time and Distance for a Single Trajectory.} (a) The travel time and distance, as observed at the source/receiver position, are parametrically plotted as a function of Mino time, $s$, for the longest trajectory, $\rsr = 10.61$. The results are  use to quantify the evolving perceived speed of light, shown in panel (b). The slope of the red line is the actual speed of light, and the discrepancy between this reference and the green curve is the perceived local slowing of light. Results are given in both geometrized units of length and, with $M = 10 M_\odot$, SI units. $a = 0.99$.} 
	\label{Dilation_1}
\end{figure}

Eqs. \ref{traj1}$_{1-3}$ allow this path length to be described analytically, and Eq. \ref{traj1}$_4$ provides the associated temporal change. These can be parametrically plotted, as in Fig. \ref{Dilation_1}(a). The slope of the (green) curve gives the speed of light, as perceived from an observer in a distant region of flat space. The slope of the red line is the actual speed of light, and the discrepancy between these quantifies the perceived local slowing of light. The left-side scaling gives the arc length in units of geometrized black hole mass, while the right-side scaling provides the SI equivalent under the assumption that the black hole is of 10 solar masses. Analogous scaling is provided at bottom and top for the elapsed Boyer-Lindquist time.  It is also possible to plot the local speed explicitly, as in Fig. \ref{Dilation_1}(b). The left-side scaling gives the perceived fraction of the speed of light, while the right-side scaling provides the SI equivalent, again assuming 10 solar masses for the SI conversion. As expected, the perceived speed of light slows as the ergosphere is approached.

%
%

\begin{figure}[t]
	\begin{center}
		\includegraphics[width=0.8\linewidth]{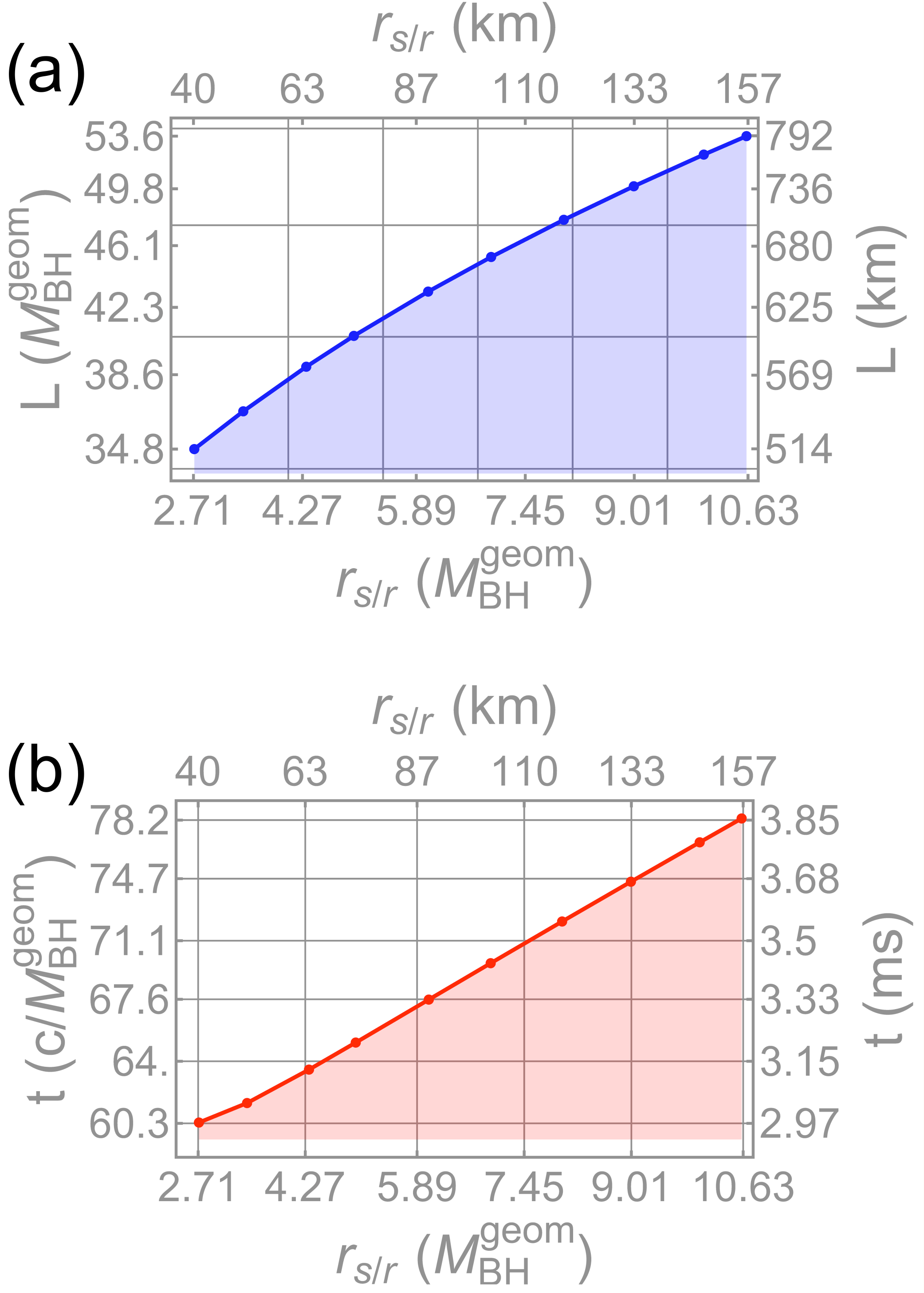}
	\end{center}
	\caption{ \emph{Total Time and Distance for All Trajectories.}  For each trajectory, the travel time and distance observed at the source/receiver position can be calculated. Results are given in both geometrized units of length and SI units and, with $M = 10 M_\odot$, SI units. $a = 0.99$. Solid curves are a guide to the eye.} 
	\label{Dilation_2}
\end{figure}

Trends in space and time dilation are quantified in Fig. \ref{Dilation_2}, where the total path length and total travel time are provided as functions of the source/receiver position. Over the entire range of closed circuits, the path length changes by a factor of 1.54, while the elapsed time changes by a factor of 1.30. While important to account for, the travel time does not undergo any excursions in association with even the closest brush across the ergosphere surface. This is because it is the 4-dimensional spacetime metric that is used to determine dilations---i.e. the convenient (3+1) foliation of spacetime does not influence the dilation analysis, and the approach could be used to quantify dilations for trajectories that transit the ergosphere as well.


%

\subsection{Trajectory Closure in Spacetime}

\begin{figure}[t]
	\begin{center}
		\includegraphics[width=0.7\linewidth]{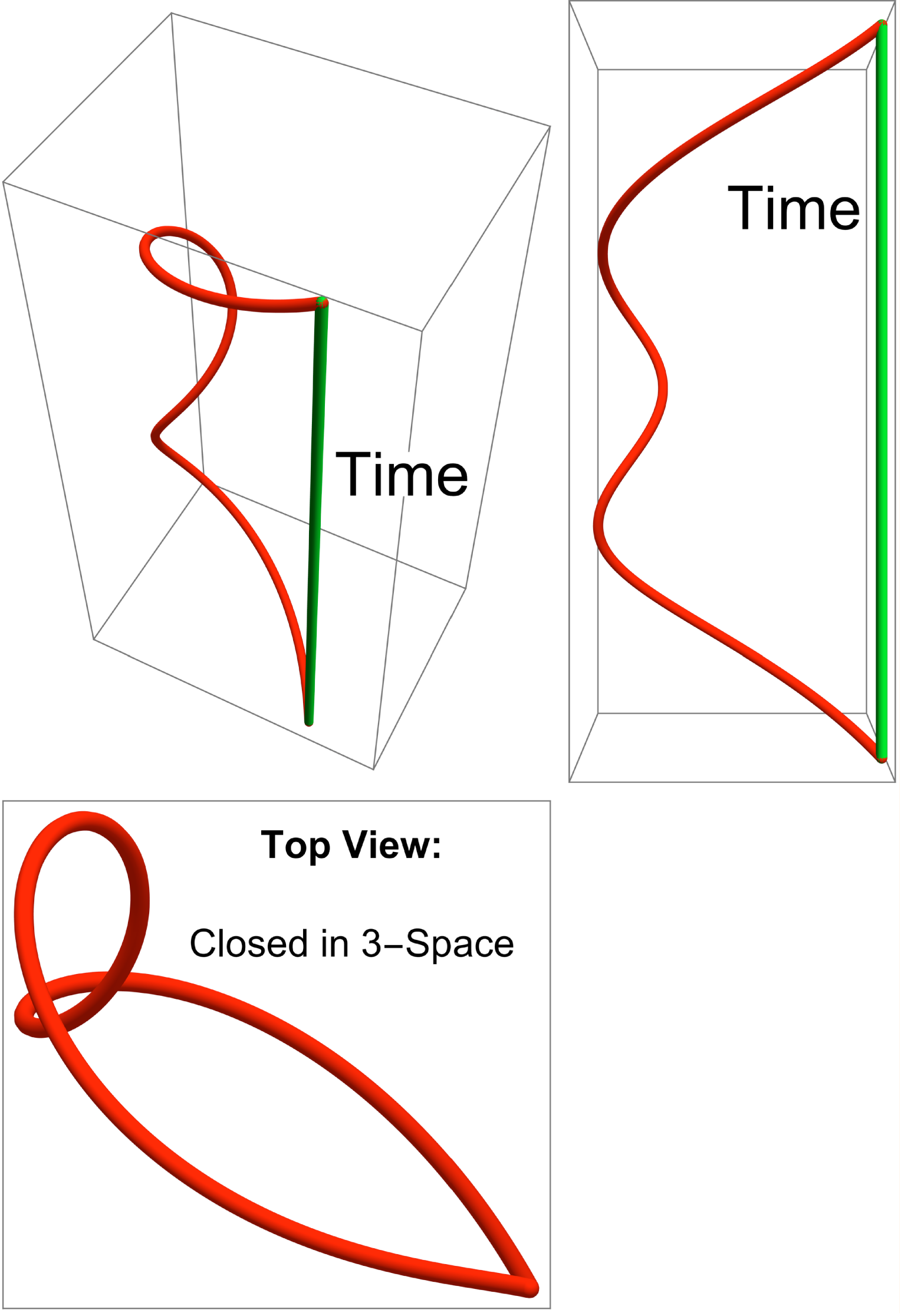}
	\end{center}
	\caption{ \emph{Trajectory Closure in Spacetime.} A cartoon is shown in (2+1) space in which an optical path (red) is closed in 2-space but not in the associated spacetime. Closure is then provided by a second source of light (green), which plays an ancillary role as a reference that moves forward in time but does not accumulate any GFR.} 
	\label{Trajectory_Closure_Spacetime}
\end{figure}

Engineered trajectories are closed in three-space but not in spacetime, and this might seem contrary to the notion of a four-space holonomy. The remedy is to consider two sources of light, with one that completes the original closed circuit (red in Fig. \ref{Trajectory_Closure_Spacetime} while the other stays in the lab, serving an ancillary role as a reference that moves forward in time (green in Fig. \ref{Trajectory_Closure_Spacetime}). The combination of these two spacetime paths comprises a closed trajectory in four-space, in the Aharonov-Bohm sense\cite{Aharonov_1959, Healey_1997}, and for which there is no Faraday rotation along the ancillary route. This provides a means of interpreting the measured GFH as a holonomy on the spacetime albeit with the understanding that the measure itself requires projection. 

%
%

\section{Conclusions}

The path of light in Kerr spacetime is fully characterized by an initial position and conserved scalars associated with metric symmetries in time, azimuth, and the Killing-Yano tensor. These appear as scalar ratios in a two-dimensional parameter space that can be demarcated into regions based on trajectory type---e.g. scattering or inward spiraling. In the present work, we add to this understanding with the discovery of a curved one-dimension interval in the parameter space that corresponds to non-equatorial, asymmetric, closed trajectories in spacetime. These have a single radial turning point midway, where they dive through the equatorial plane near to the ergosphere horizon. The parameter interval is bounded at one end by a trajectory that enters the ergosphere and at the other by a trajectory that is iso-radial and highly symmetric.

As is true for most non-equatorial trajectories, the propagation of linearly polarized light along such ambits produces a gravitational Faraday rotation, a three-space perception that the plane of polarization is not simply twisting in response to path curvature. What is striking about these circuits, though, is that the rotation exhibits holonomy due to path asymmetry. 

Interestingly, this gravitational Faraday holonomy (GFH) increases as the source/receiver is moved outwards within the equatorial plane. In fact, the rate of accumulation of GFR is only large over a small region near the radial turning point. As the source is moved further out, a larger gravitational slingshot effect is required to swing it back towards the point of origin. The turning point thus move inwards towards the ergosphere. As a result, the spacetime encountered is more distorted, manifested as a polar, diagonal component of the three-space metric component that becomes ever larger. This, in turn, can be traced back to the temporal, diagonal component of the four-space metric, which decreases towards zero as the ergosphere is approached. It is therefore a severe, localized time dilation that underlies the rotation rate excursion and concomitant trend of an increasingly large gravitational holonomy as the source is moved outward.

The well-understood and quantified trends in holonomy accumulation with source/receiver position and black hole rotation rate make it possible to envision experiments in which gravitational Faraday holonomy can be experimentally measured. This has a photonic extension as well that could be adapted to study the effect of entanglement on holonomy. The setting also lends itself to a study of the Spin-Hall effect of light. 

\appendix

\section{Equations of Motion}
Closed-form expressions for the trajectory components can be constructed that satisfy the Equations of Motion, Eqs. \ref{EoMr}--\ref{EoMt}. The process is tedious, but the ultimate result, Eqs. \ref{traj1}, is very practical and easy to use. Only the essential equations are provided below, since a detailed development is given elsewhere\cite{Gralla_2020}. The expressions given for $\phi(s)$ and $t(s)$  differ from that earlier analysis, though, as there seems to be a number of small errors in that earlier work.  In particular, radial functions $F_2(r)$, $E_2(r)$, $\Pi_1(r)$, and $\Pi_\pm(r)$ have been modified as have Mino time functions $I_\pm (s)$, $I_j (s)$ with $j = 0, 1, 2$, and $I_t(s)$. These corrected analytical functions give results essentially identical to numerical solutions of the equations of motion for a wide range of initial conditions and choices of scalar invariants. See further discussion of this associated with the analytical/numerical comparison of Fig. \ref{NumericalChecks}.
\vskip 0.5 cm
\noindent Define constants $A$, $B$, and $C$.
\begin{align}
A &:= a^2 - \eta - \lambda^2 \nonumber \\
B &:= 2(\eta + (\lambda - a)^2) \nonumber \\
C &:= -a^2 \eta
\end{align}
Define constants $P$, $Q$, $w_\pm$ and $z$.
\begin{align}
P &:= -\frac{A^2}{12} - C \nonumber \\
Q &:= -\frac{A}{3} \left(  \left( \frac{A}{6} \right)^2 - C \right) - \frac{B^2}{8} \\
w_\pm &:= \left( \pm\sqrt{\frac{P^3}{27}+\frac{Q^2 }{4}}-\frac{Q}{2} \right)^{1/3}  \nonumber \\
z &:= \sqrt{\frac{w_+ + w_-}{2} - \frac{A}{6} } \nonumber 
\end{align}
Define roots of the radial potential, $\cal R$,  $r_1$, $r_2$, $r_3$, and $r_4$, as well as radii of the inner and outer event horizons, $r_\pm$.
\begin{align}
r_1 &:= -\sqrt{-\frac{A}{2}+\frac{B}{4
   z}-z^2} - z \nonumber \\
r_2 &:= \sqrt{-\frac{A}{2}+\frac{B}{4
   z}-z^2} - z \nonumber \\
r_3 &:=   -\sqrt{-\frac{A}{2}+\frac{B}{4
   z}-z^2} + z  \\
r_4 &:=   \sqrt{-\frac{A}{2}+\frac{B}{4
   z}-z^2} + z \nonumber \\
 r_\pm &:=   1 \pm \sqrt{1-a^2}\nonumber 
 \end{align}

Define radial function $x_2(r)$ and constant $k_2$.
\begin{align}
x_2(r) &:= \sqrt{\left( \frac{r - r_4}{r - r_3} \right) \left( \frac{r_3 - r_1}{r_4 - r_1}\right) } \nonumber \\
k_2 &:=\frac{\left(r_3-r_2\right)
   \left(r_4-r_1\right)}{\left
   (r_3-r_1\right)
   \left(r_4-r_2\right)}
\end{align}
Define constants $\Delta_\theta$ and $u_\pm$.
\begin{align}
\Delta_\theta &:= \frac{1}{2} \left( 1 - \frac{\eta + \lambda^2}{a^2} \right) \nonumber \\
u_\pm &:= \Delta_\theta  \pm \sqrt{\Delta_\theta^2 + \eta/a^2} 
\end{align}
Define radial functions $F_2(r)$, $E_2(r)$, $\Pi_1(r)$, and $\Pi_\pm(r)$. These are expressed in terms of Jacobi elliptic functions $E$, $F$, and $\Pi$.
\begin{align}
F_2(r) &:= \frac{2 F\left(\sin^{-1}\left(x_2(r)\right) , k_1\right)}{\sqrt{\left(r_3-r_1\right) \left(r_4-r_2\right)}} \nonumber \\
 E_2(r) &:= \sqrt{\left(r_3 - r_1\right)  \left(r_4 - r_2\right) } E\left(\sin^{-1}\left(x_2(r)\right) , k_2\right) \nonumber \\
 \Pi_1(r) &:= \frac{2 \Pi\left( \frac{r_4 - r_1}{r_3 - r_1}, \sin^{-1}\left(x_2(r)\right) , k_2\right)}{\sqrt{\left(r_3-r_1\right) \left(r_4-r_2\right)}} \\
  \Pi_\pm(r) &:= \frac{2 }{\sqrt{\left(r_3-r_1\right) \left(r_4-r_2\right)}} \frac{r_4 - r_3}{(r_\pm - r_3)(r_\pm - r_4)}  \nonumber \\ 
& \quad \times\Pi\left[ \left( \frac{r_\pm - r_3}{r_\pm - r_4}\right) \left( \frac{r_4 - r_1}{r_3 - r_1}\right) , \sin^{-1}\left(x_2(r)\right) , k_2\right] \nonumber 
\end{align}
Define radial functions $\mathcal{I}_{0}(r)$, $\mathcal{I}_{1}(r)$, $\mathcal{I}_{2}(r)$, and $\mathcal{I}_{\pm}(r)$. 
\begin{align}
\mathcal{I}_{0}(r) &:= F_2(r) \nonumber \\
\mathcal{I}_{1}(r) &:= r_3 F_2(r) + (r_4 - r_3) \Pi_1(r) \nonumber \\
\mathcal{I}_{2}(r) &:= \frac{\sqrt{\mathcal{R}(r)}}{r - r_3} -  \frac{1}{2} (r_1 r_4 + r_2 r_3) F_2(r) - E_2(r)   \\
\mathcal{I}_{\pm}(r) &:=-\Pi_\pm(r) -\frac{F_2(r)}{r_\pm - r_3} \nonumber  
\end{align}
Define Mino time function $X_2(s)$.
\begin{equation}
X_2(s) := \frac{1}{2}\sqrt{(r_3 - r_1)(r_4 - r_2)} \bigl(s + \nu_r  F_2(r_{\scriptstyle init}) \bigr) 
\end{equation}
Define constants $\mathcal{G}_\theta$ and $\nu_\theta$.
\begin{align}
\mathcal{G}_\theta &:=\frac {-1} {\sqrt{-a^2 u_-}} F\left( \sin^{-1}\left( \frac{\cos \theta_{\scriptstyle init}}{\sqrt{u_+}} \right) ,\frac{u_+}{u_-}\right) \\
\nu_\theta &:= {\rm Sign(p^3_{\scriptstyle init})} \equiv {\rm Sign(p^\theta_{\scriptstyle init})} \nonumber 
\end{align}
Define constant $\mathcal{G}_\phi$ and Mino time function $G_\phi(s)$, where the latter is expressed in terms of the Jacobi amplitude function, ${\rm am}$.
\begin{align}
\mathcal{G}_\phi &:=\frac {-1} {\sqrt{-a^2 u_-}} \Pi \left[ u_+,\sin^{-1}\left( \frac{\cos \theta_{\scriptstyle init}}{\sqrt{u_+}} \right) ,\frac{u_+}{u_-}\right] \nonumber \\
G_\phi(s) &:=  - \nu_\theta \mathcal{G}_\phi \\
+&\frac {1}{\sqrt{-a^2 u_-}}  \Pi \left[ u_+, {\rm am} \left( \sqrt{-a^2 u_-} (s + \nu_\theta \mathcal{G}_\theta),  \frac{u_+}{u_-} \right) ,  \frac{u_+}{u_-} \right]   \nonumber 
\end{align}
Define Mino time functions $I_\pm (s)$.
\begin{align}
I_\pm (s) &:= -\nu_r \sign\bigl(F_2(r_{\scriptstyle init}) - s\bigr) \left( \Pi_\pm(r(s))  + \frac{F_2(r_{\scriptstyle init})}{r_\pm - r_3}\right) \nonumber \\ 
& - \nu_r \mathcal{I}_{\pm}(r_{\scriptstyle init}) 
\end{align}
Define Mino time functions $I_j (s)$ with $j = 0, 1, 2$.
\begin{equation}
I_j (s) := \nu_r \left[ \sign\bigl(F_2(r_{\scriptstyle init}) - s\bigr) \mathcal{I}_j (r_{\scriptstyle init}) - \mathcal{I}_j (r_{\scriptstyle final} ) \right] 
\end{equation}
Define Mino time function $I_t(s)$.
\begin{align}
I_t(s) &:= \frac{4}{r_+ - r_-} \left[ r_+\left( r_+ - \frac{a \lambda}{2} \right) I_+(s)  \right] \nonumber \\ 
&\,\, - \frac{4}{r_+ - r_-} \left[ r_-\left( r_- - \frac{a \lambda}{2} \right) I_-(s)  \right]  \\
& \,\, + 4 I_0(s)   + 2 I_1(s) + I_2(s) \nonumber 
\end{align}
Define Mino time function $G_t(s)$ in terms of Jacobi elliptic functions, $E$ and $F$, as well as the Jacobi amplitude function, ${\rm am}$.
%
\begin{align}
G_t(s) &:= \frac{-u_+}{\sqrt{-a^2 u_-}}  \left( \frac{u_-}{u_+} \right)  \nonumber \\
&\quad \times E\left[ \rm{am}\left( \sqrt{-a^2 u_-} (s + \nu_\theta\mathcal{G}_\theta), \frac{u_+}{u_-} \right), \frac{u_+}{u_-} \right]  \nonumber \\
& \,\,+ \frac{u_+}{\sqrt{-a^2 u_-}}  \left( \frac{u_-}{u_+} \right) \\
&\quad \times F\left[ \rm{am}\left( \sqrt{-a^2 u_-} (s + \nu_\theta\mathcal{G}_\theta), \frac{u_+}{u_-} \right), \frac{u_+}{u_-} \right]  \nonumber 
\end{align}

Using the plethora of support functions given above, trajectory components can now be defined as functions of Mino time, $s$.
\begin{align}\label{traj2}
r(s) &= \frac{r_4 (r_3 - r_1) - r_3(r_4 - r_1){\rm sn}^2(X_2(s), k)}{(r_3 - r_1) - (r_4 - r_1){\rm sn}^2(X_2(s), k) } \nonumber \\
\theta(s) &= \cos^{-1} \left[ -\nu_\theta \sqrt{u_+} \,\,{\rm sn} \left(  \sqrt{-a^2 u_-} (s + \nu_\theta \mathcal{G}_\theta ), \frac{u_+}{u_-} \right) \right] \nonumber \\
\phi(s) &= \lambda G_\phi(s)  \nonumber\\
 + &\frac{2 a}{r_+ - r_-} \left[ \left( r_+ - \frac{a \lambda}{2} \right) I_+(s) - \left( r_- - \frac{a \lambda}{2} \right) I_-(s) \right] \nonumber \\
t(s) &= I_t(s) + a^2 G_t(s) . 
\end{align}
%


\end{document}